\documentclass[sigconf]{acmart}

\usepackage{float}
\usepackage{amsmath,amsfonts,mathtools}
\usepackage{algpseudocode}
\usepackage{array,multirow,graphicx}
\usepackage{textcomp}
\usepackage{enumitem}
\usepackage{threeparttable}
\usepackage[multiple]{footmisc}
\usepackage{xcolor,colortbl,soul}
\usepackage{booktabs} %

\usepackage{graphicx}
\usepackage{subcaption}
\usepackage[english,ruled,vlined, slide, norelsize]{algorithm2e}
\usepackage{hyperref}

\AtBeginDocument{%
  }

\newcommand*{\LastTimeAccessed}{, Accessed: March 2024}

\copyrightyear{2024}
\acmYear{2024}
\setcopyright{rightsretained}
\acmConference[ICPE '24]{Proceedings of the 15th ACM/SPEC International Conference on Performance Engineering}{May 7--11, 2024}{London, United Kingdom}
\acmBooktitle{Proceedings of the 15th ACM/SPEC International Conference on Performance Engineering (ICPE '24), May 7--11, 2024, London, United Kingdom}\acmDOI{10.1145/3629526.3645042}
\acmISBN{979-8-4007-0444-4/24/05}

\begin{document}

\title{Daedalus: Self-Adaptive Horizontal Autoscaling for Resource Efficiency of Distributed Stream Processing Systems}

\author{Benjamin J. J. Pfister}
\affiliation{
  \institution{Technische Universität Berlin}
  \city{Berlin}
  \country{Germany}
}
\email{benjamin.pfister@campus.tu-berlin.de}

\author{Dominik Scheinert}
\affiliation{
  \institution{Technische Universität Berlin}
  \city{Berlin}
  \country{Germany}
}
\email{dominik.scheinert@tu-berlin.de}

\author{Morgan K. Geldenhuys}
\affiliation{
  \institution{Technische Universität Berlin}
  \city{Berlin}
  \country{Germany}
}
\email{morgan.geldenhuys@tu-berlin.de}

\author{Odej Kao}
\affiliation{
  \institution{Technische Universität Berlin}
  \city{Berlin}
  \country{Germany}
}
\email{odej.kao@tu-berlin.de}

\begin{abstract}
Distributed Stream Processing (DSP) systems are capable of processing large streams of unbounded data, offering high throughput and low latencies.
To maintain a stable Quality of Service (QoS), these systems require a sufficient allocation of resources.
At the same time, over-provisioning can result in wasted energy and high operating costs.
Therefore, to maximize resource utilization, autoscaling methods have been proposed that aim to efficiently match the resource allocation with the incoming workload.
However, determining when and by how much to scale remains a significant challenge.
Given the long-running nature of DSP jobs, scaling actions need to be executed at runtime, and to maintain a good QoS, they should be both accurate and infrequent.
To address the challenges of autoscaling, the concept of self-adaptive systems is particularly fitting. 
These systems monitor themselves and their environment, adapting to changes with minimal need for expert involvement.

This paper introduces \emph{Daedalus}, a self-adaptive manager for autoscaling in DSP systems, which draws on the principles of self-adaption to address the challenge of efficient autoscaling.
Daedalus monitors a running DSP job and builds performance models, aiming to predict the maximum processing capacity at different scale-outs.
When combined with time series forecasting to predict future workloads, Daedalus proactively scales DSP jobs, optimizing for maximum throughput and minimizing both latencies and resource usage.
We conducted experiments using Apache Flink and Kafka Streams to evaluate the performance of Daedalus against two state-of-the-art approaches.
Daedalus was able to achieve comparable latencies while reducing resource usage by up to 71\%.

\end{abstract}



\maketitle

\section{Introduction}

Distributed Stream Processing (DSP) is an important paradigm that enables quick extraction of insights from unbounded data streams with high throughput and low latencies.
The generation of streaming data is continually increasing and this trend is evident across a range of contexts, including online advertising, financial transactions, and IoT sensor networks~\cite{Cardellini2022RuntimeAO,Vogel2022SelfadaptationOP,OpenBenchmark2020}.
In order to provide a good Quality of Service (QoS), DSP systems need to be properly configured. 
Insufficient allocation of resources can lead to unstable service delivery, while over-provisioning results in wasted energy and higher operational costs. 
Achieving the right balance in resource allocation is therefore crucial to optimize both performance and cost-effectiveness.
However, due to the dynamic nature of streaming workloads, configurations can quickly become obsolete.
Likewise, because the manual tuning of configurations is infeasible over the course of a long-running job, resources can only be adjusted through an automated approach.
Therefore, it is prudent to provide DSP systems with self-adaptive capabilities, enabling them to monitor themselves and respond to environmental changes by autonomously tuning their configurations during runtime.

Automated scaling of computational resources, commonly known as autoscaling, serves as the key method for aligning resources with dynamic workloads in DSP systems.
The scaleout, defined by the number of worker nodes and processing slots per node, directly determines the system's level of parallelism and influences the overall resource allocation. 
This, in turn, affects each worker node's processing potential. 
Popular DSP frameworks like Apache Flink \cite{Flink2015}, Kafka Streams \cite{Sax2018StreamsAT}, and Apache Spark \cite{Spark2016}, support dynamic adjustments in scaleout during runtime, which impacts the job's parallelism and subsequently its resource utilization. 
Typically, these systems are deployed in cloud environments, where resources are provisioned elastically and can be scaled in or out as required, directly affecting the processing capabilities of the DSP job.

Autoscaling approaches are optimized towards various adaptation goals, with most research primarily focusing on enhancing DSP system performance by maximizing throughput or minimizing latency~\cite{Cardellini2022RuntimeAO, Rger2019ACS}. 
Additionally, secondary goals like minimizing resource usage, reducing monetary costs, lowering energy consumption, or shortening recovery times are also considered~\cite{Cardellini2022RuntimeAO, Vogel2022SelfadaptationOP}. 
Despite active research in this area, there is still potential for further advancements. 
Many existing methods demand in-depth system or job knowledge, such as setting scaling thresholds or altering DSP source code~\cite{Kalavri2018ThreeSI,Kalim2019CaladriusAP,Zhang2021AuTraScaleAA}. 
Often, scaling decisions lead to downtime for initializing new workers or recalculating data distribution among operators, yet few strategies account for this overhead. 
Additionally, many presume an even distribution of data across parallel operators, which is not always the case in practical scenarios.

This paper approaches the challenges of autoscaling DSP systems from the perspective of self-adaptation. 
It proposes a self-adaptive manager called Daedalus that targets a running DSP job and horizontally scales its parallelism to adapt to the incoming workload while minimizing resource usage. 
In order to meet QoS requirements, Daedalus can optimize towards a target recovery time and ensure that a job will recover between scaling actions.
Because scaling decisions incur an overhead cost, Daedalus employs Time Series Forecasting (TSF) to predict the future workload in order to reduce the frequency of scaling actions. 
By ensuring that the incoming workload can be processed and minimizing system downtime, Daedalus can achieve reasonable latencies. 
Scaling decisions are realized through a combination of monitoring, performance modeling, and TSF. 
Unlike most existing approaches that ignore how data is split among parallel workers, Daedalus explicitly incorporates data skew in its capacity models. 
It is a general approach, applicable to containerized DSP systems running in cloud environments. 
Daedalus has been evaluated using three benchmark DSP jobs, employing two DSP frameworks, Apache Flink and Kafka Streams, and is compared against two state-of-the-art approaches.

This paper is structured as follows: 
\autoref{sec:related_work} reviews related work on autoscaling, recovery time, and time series forecasting.
\autoref{sec:approach} describes the approach taken to realize self-adaptive autoscaling. 
In \autoref{sec:evaluation}, the approach is evaluated, and the results are discussed. 
Lastly, the paper is concluded in \autoref{sec:conclusion}.
\section{Related Work}
\label{sec:related_work}

Adaptive autoscaling for DSP systems is an active field of research. 
This section describes the most relevant state-of-the-art solutions that Daedalus builds upon and work relevant to recovery time and TSF.

Heinze et al. \cite{Heinze2014AutoscalingTF} assess the viability of both threshold-based and reinforcement learning autoscaling techniques using their approach called FUGU. The authors find that reinforcement learning produces scaling decisions that best maximize system utilization, and that global thresholds are not well-suited for autoscaling. However, it should also be noted that reinforcement learning techniques can take a long time to adequately train and need to make ill-suited scaling decisions to learn.

Dhalion \cite{Floratou2017DhalionSS} is a self-adaptive system developed for Heron that employs user-defined policies to enable autoscaling. By monitoring metrics such as backpressure, tuples waiting in buffers, and data skew across operators, Dhalion determines if resources are over-provisioned or under-provisioned and scales in or out accordingly. Although Dhalion aims to be a comprehensive self-adaptive DSP solution, defining suitable policies requires expert knowledge of the job and system. Additionally, its autoscaling capabilities have been found to over-provision resources. Because individual operators are scaled one at a time after detecting backpressure, Dhalion needs a long time to converge to an optimal configuration \cite{Kalavri2018ThreeSI}.

DS2 is a reactive autoscaling approach that monitors a running DSP job to calculate true processing rates in order to determine the proportional processing relationships among operators \cite{Kalavri2018ThreeSI}. 
By scaling in response to the workload, DS2 can accurately adjust the parallelism of operators. However, DS2 assumes data skew is not present and that workloads remain stable during scaling operations. It also requires manual implementation of the true processing rate metric, as this metric is not readily available in all DSP systems.

AuTraScale \cite{Zhang2021AuTraScaleAA} utilizes Bayesian Optimization for operator autoscaling to minimize latency while maximizing total throughput. Like DS2, it uses the true processing rates to find the minimum parallelism needed to process a static workload. When the system is over-provisioned or latency exceeds a target threshold, AuTraScale reactively rescales. Although designed to find the optimal configuration for a static workload, AuTraScale includes a transfer learning algorithm to more quickly find optimal configurations if the input rate would change. However, the authors do not evaluate their approach on a dynamic workload and assume no data skew.

Caladrius \cite{Kalim2019CaladriusAP} is a performance modeling tool that uses TSF to anticipate future workloads and predict the throughput and CPU usage for both the current and other scale-outs to enable proactive scaling decisions. Like DS2, Caladrius models the operator topology and relationship between operator input and output rates. The authors make the important observation that throughput rates and CPU utilization are linearly related. Therefore, by predicting the throughput, the authors can predict CPU utilization. Daedaulus is heavily influenced by this and uses the linear relationship between throughput and CPU to estimate capacity. 
However, whereas Daedalus uses a CPU-throughput regression model to estimate capacity, Caladrius relies heavily on backpressure to observe maximum throughput rates, which is an unreliable metric in presence of data skew or slow nodes.

Relatively few DSP autoscaling approaches incorporate the overhead cost of scaling decisions \cite{Vogel2022SelfadaptationOP}. Phoebe chooses scale-outs that can guarantee a target recovery time \cite{Geldenhuys2022PhoebeQD}. Martin et al. \cite{Martin2015UserConstraintAS} provide a self-adaptive approach for a DSP system to adjust its fault tolerance mechanism during runtime. While not an autoscaling approach, their approach allows users to provide high level constraints such as a target recovery time. Lastly, Borkowski, Hochreiner, and Schulte \cite{Borkowski2019MinimizingCB} note the downtime caused by autoscaling and aim to reduce the number of rescaling actions for a threshold-based autoscaler. Their approach uses an extended Kalman filter to estimate trends in the workload, similar to ARIMA. By ignoring short-term variations in the workload, they are able to reduce the number of scaling decisions, thereby reducing overall recovery time over the course of the DSP job.

Gontarska et al. \cite{Gontarska2021EvaluationOL} compare commonly used TSF methods to assess their use in predicting DSP workloads. They compare seven methods, including ARIMA and two deep learning methods. Although deep learning methods produced the best predictions overall, they also required much longer training times. The ARIMA model, on the other hand, was trained faster and yielded good results for making short-term predictions in the tested 5 and 15 minute forecasts. Caladrius and Phoebe both use TSF to predict future workloads \cite{Kalim2019CaladriusAP,Geldenhuys2022PhoebeQD}. Caladrius uses Facebook's Prophet while Phoebe also uses ARIMA.

\section{Approach}
\label{sec:approach}

\begin{figure}
    \centering
   \includegraphics[width=\linewidth]{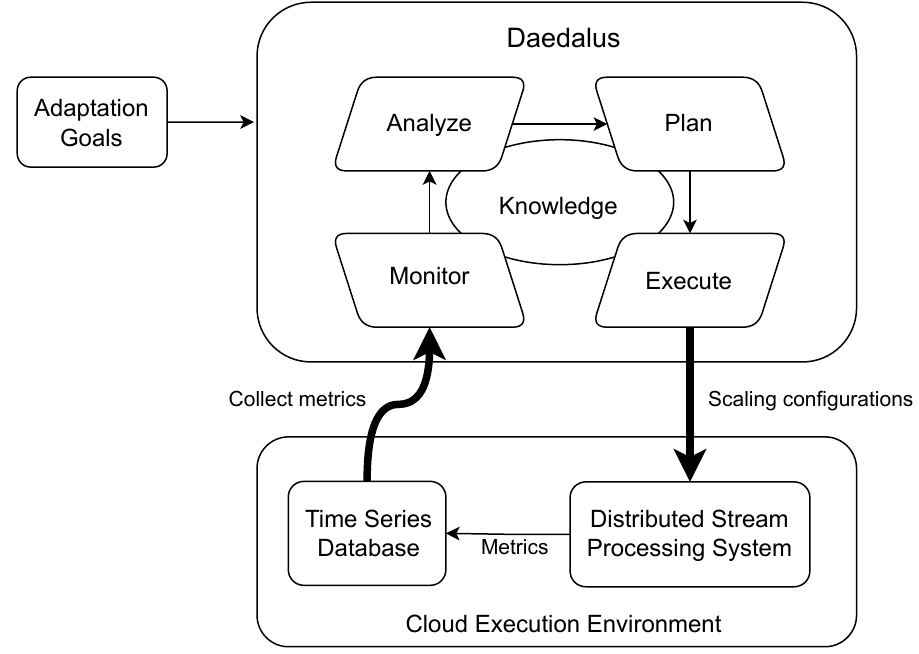}
   \caption{Daedalus architecture overview}
   \label{fig:daedalus-architecture}
\end{figure}

In this paper, the challenges of autoscaling a DSP system for performance optimization are approached from the perspective of self-adaptation~\cite{DBLP:conf/wosp/PorterMGA18,DBLP:conf/wosp/0001BWY18,DBLP:conf/wosp/Perez-PalacinM14,DBLP:conf/wosp/EwingM14}.
Using self-adaptive autoscaling enables a DSP system to monitor itself and make scaling decisions to meet DSP requirements and adaptation goals while processing dynamic workloads. This paper proposes a self-adaptive DSP autoscaler called Daedalus. Its main objectives are to ensure enough resources are allocated to process the incoming workload while minimizing resource usage, meet QoS requirements by targeting a recovery time, and providing a stable level of service by enacting long-lived scaling decisions over the course of a long-running DSP job.
On a high level, Daedalus uses the self-adaptive MAPE-K control loop to continuously monitor a running DSP job by collecting metrics stored in a time series database. It analyzes the data, builds models to estimate the capacity across all potential scale-outs, and predicts the future workload using TSF. Using both the historical and predicted workload, it decides if a scaling action is necessary and determines how to scale based on the described adaptation goals and estimated recovery time. Lastly, it executes a scaling action if necessary. A high-level overview of the approach, its architecture, and components can be seen in~\autoref{fig:daedalus-architecture}.

In order to determine when and how to scale, Daedalus builds capacity models at the worker level using throughput and CPU utilization metrics. These metrics are typically already exposed by DSP systems so that their performance can be monitored. Monitoring provides accurate, up-to-date insights for a job running in a cloud environment. Previous observations, including a series of profiling runs conducted at the beginning of a deployment as in~\cite{Geldenhuys2022PhoebeQD}, have the potential to become less reliable over time. Since the underlying resources or placement of operators can change over the course of a long-running job, continually collecting metrics and monitoring the system performance is prudent.

Developing separate self-adaptive autoscaling solutions for each individual DSP system and job is complex and time-intensive. Therefore, Daedalus aims to provide a general solution for containerized DSP systems by requiring only a few commonly exposed metrics. In this paper, Daedalus is tested with Flink and Kafka Streams, though it is applicable to other systems as well.

Our approach makes the assumption that the DSP system is running in a cloud environment where homogeneous resources can be elastically scaled. However, it is also taken into account that these homogeneous resources may not provide identical performance by monitoring each worker individually. 
This approach also relies on a few metrics being available. These include the throughput, CPU utilization, and consumer lag for each worker, collected from the DSP system, and the workload, collected from the data source. 

\subsection{Performance Modeling}

At the core of our approach is a capacity model, which 
provides a basis for all scaling decisions. It estimates the maximum number of tuples that can be processed per second at a given scale-out. Knowing this information along with the current and future workload informs if scaling is necessary and if so, how to scale. The capacity model is based on the observable relationship between metrics, especially, throughput and CPU utilization.

\begin{figure}
\centering
\begin{subfigure}[b]{.49\columnwidth}
  \centering
  \includegraphics[width=\textwidth]{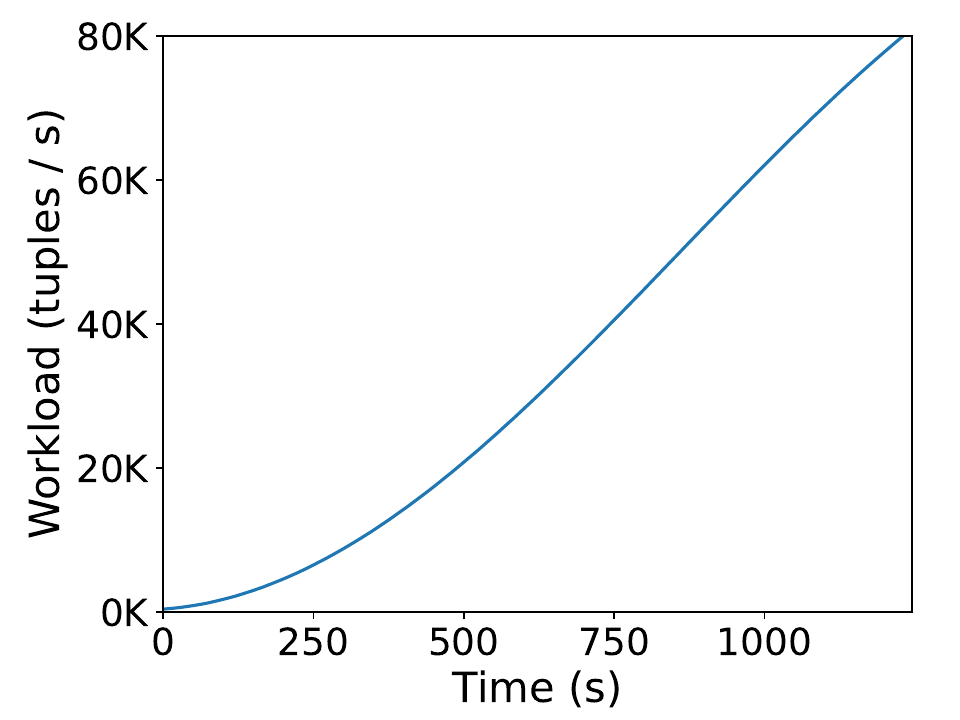}
  \caption{Workload}
  \label{fig:metric_relationships_a}
\end{subfigure}
\begin{subfigure}[b]{.49\columnwidth}
  \centering
  \includegraphics[width=\textwidth]{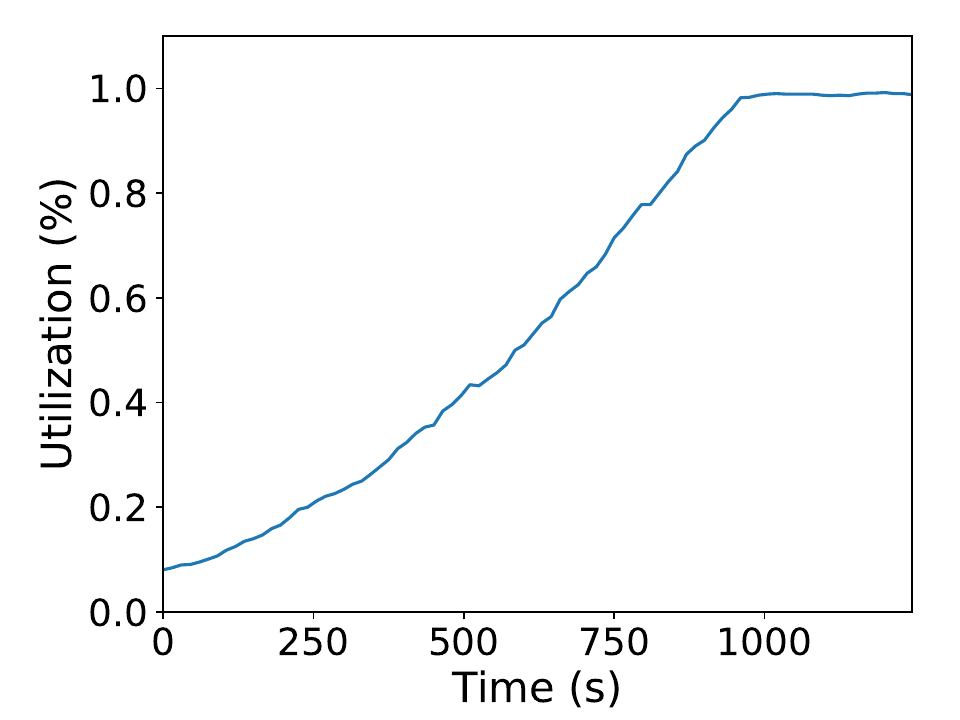}
  \caption{CPU utilization}
  \label{fig:metric_relationships_b}
\end{subfigure}
\begin{subfigure}[b]{.49\columnwidth}
  \centering
  \includegraphics[width=\textwidth]{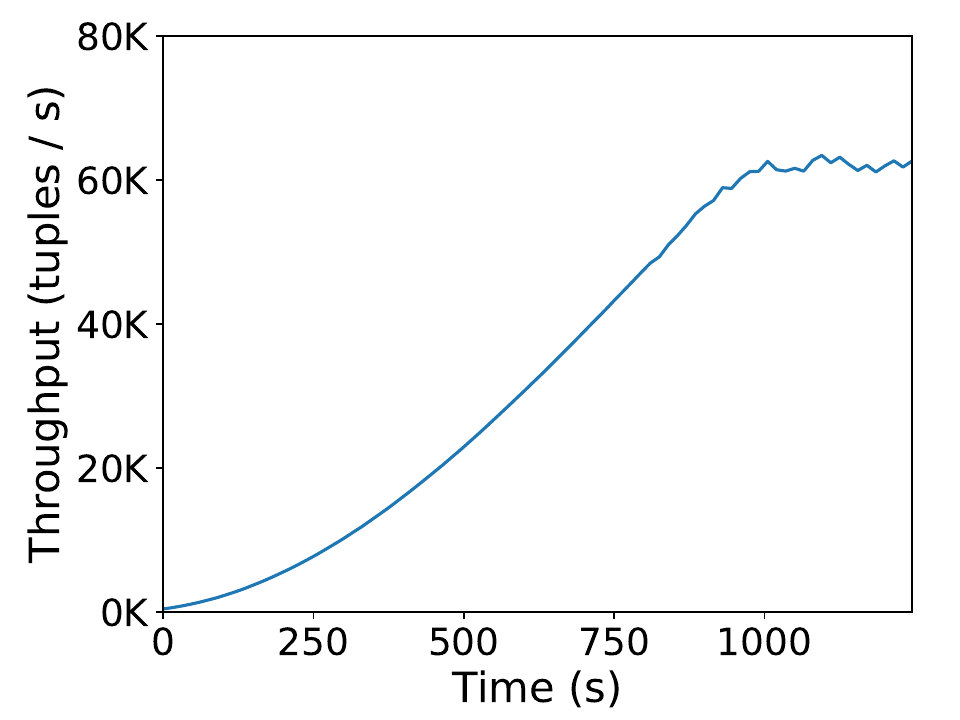}
  \caption{Throughput}
  \label{fig:metric_relationships_c}
\end{subfigure}
\begin{subfigure}[b]{.49\columnwidth}
  \centering
  \includegraphics[width=\textwidth]{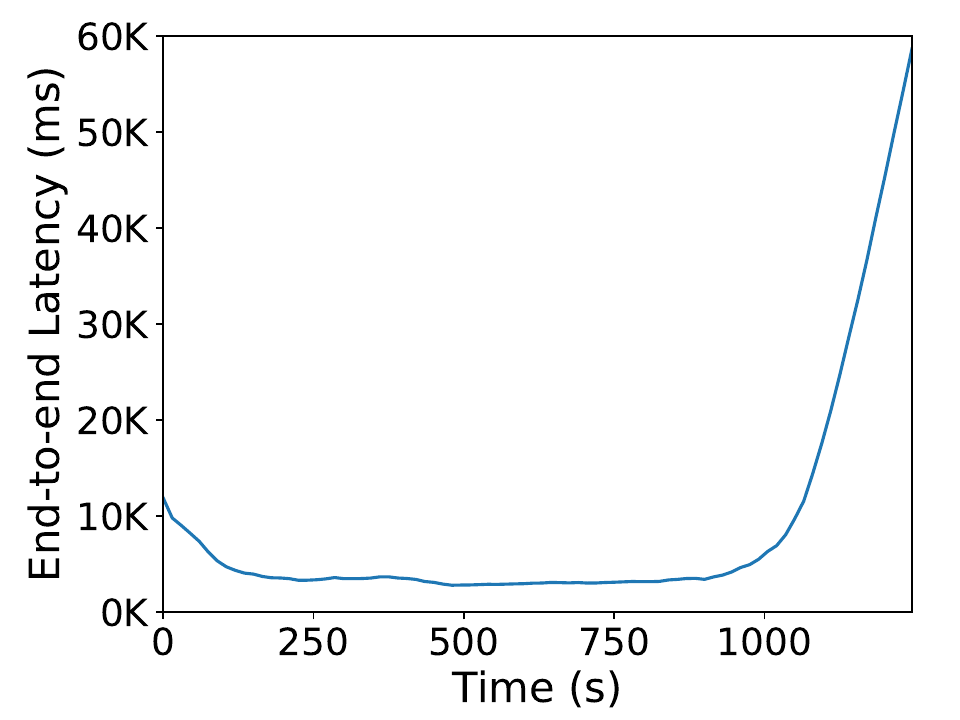}
  \caption{End-to-end latency}
  \label{fig:metric_relationships_d}
\end{subfigure}
\caption{Relationships between metrics}
\label{fig:metric_relationships}
\end{figure}

\autoref{fig:metric_relationships} shows the relationship between the incoming workload, CPU utilization, throughput, and end-to-end latency at a fixed parallelism. The metrics were taken from a running job and are therefore not influenced by a warm-up period. 
As long as sufficient resources are available, the throughput of the DSP system will match the workload.
When there are no longer resources to keep up with the rate of incoming tuples, the DSP system reaches its maximum capacity, and CPU utilization is 100\%. In this example, throughput is capped at 60,000 tuples per second.
As can be seen, the relationship between throughput and CPU utilization is linear.

End-to-end latency is impacted by several factors and is often job dependent. In jobs that use windowing, for example, end-to-end latency can increase when not enough tuples exist to trigger the end of the window. In~\autoref{fig:metric_relationships_d}, the latency is slightly influenced by the workload while processing capacity exists. However, this effect is minor when compared to the sharp increase when the workload exceeds maximum processing capacity. When the system cannot keep up with the incoming workload, either due to insufficient resources or downtime, tuples accumulate and end-to-end latency increases. It is therefore a main objective of Daedalus to ensure that sufficient capacity exists to process the incoming workload and to minimize downtime in order to limit spikes in end-to-end latency.

Much existing research on DSP autoscaling assumes that data is split equally among parallel operators, i.e. that data skew is not present. However, ignoring data skew can be a major weakness in these approaches as it is often present in reality.

\begin{figure}
\centering
\begin{subfigure}[b]{.49\columnwidth}
  \centering
  \includegraphics[width=\textwidth]{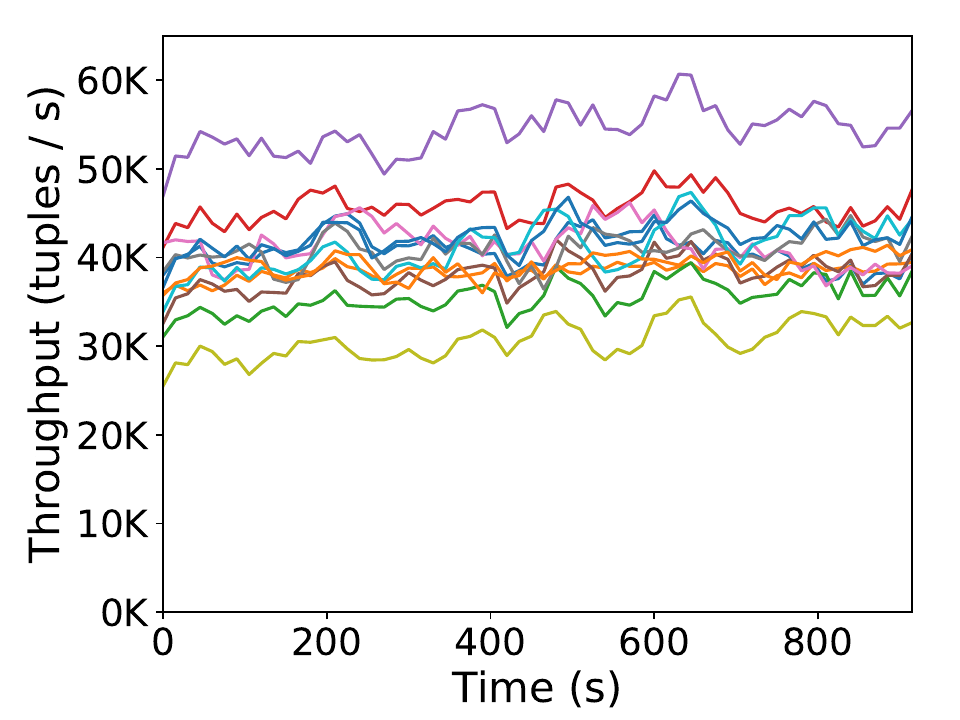}
  \caption{Throughput per worker}
  \label{fig:data_skew_a}
\end{subfigure}
\begin{subfigure}[b]{.49\columnwidth}
  \centering
  \includegraphics[width=\textwidth]{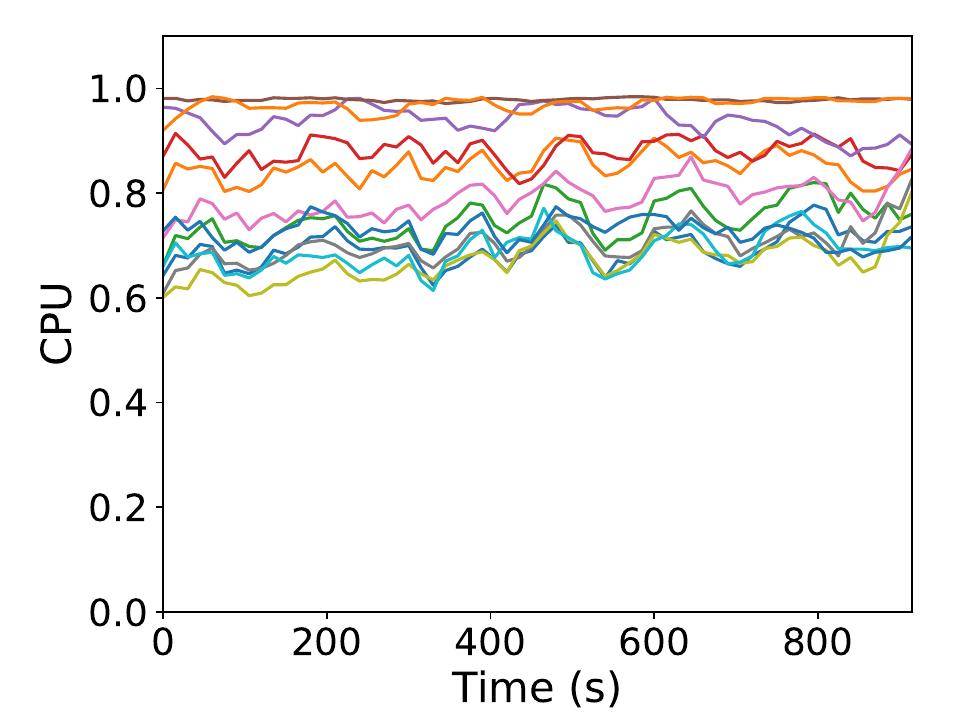}
  \caption{CPU utilization per worker}
  \label{fig:data_skew_b}
\end{subfigure}
\begin{subfigure}[b]{.49\columnwidth}
  \centering
  \includegraphics[width=\textwidth]{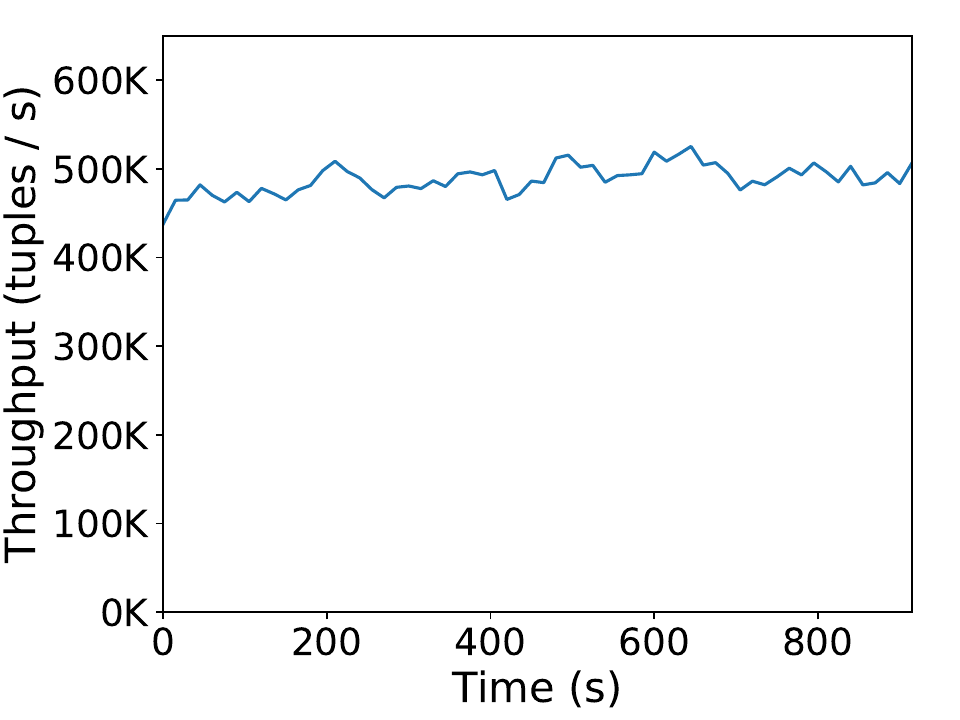}
  \caption{Total throughput}
  \label{fig:data_skew_c}
\end{subfigure}
\begin{subfigure}[b]{.49\columnwidth}
  \centering
  \includegraphics[width=\textwidth]{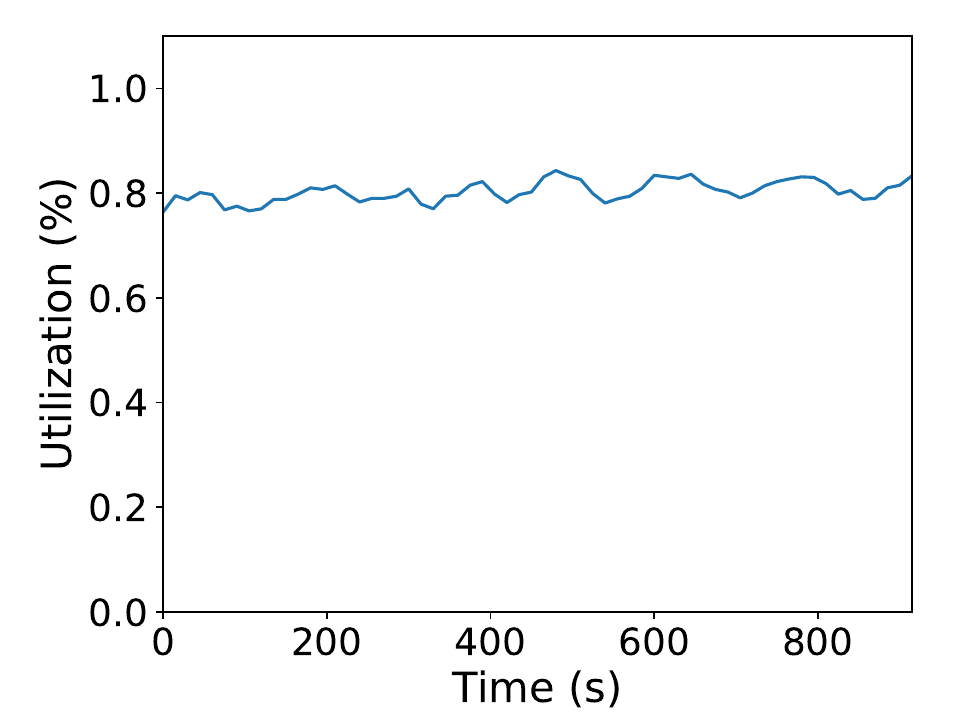}
  \caption{Average CPU utilization}
  \label{fig:data_skew_d}
\end{subfigure}
\caption[Data skew]{Maximum throughput at a parallelism of 12 showing data skew and an average CPU utilization of 0.8.}
\label{fig:data_skew}
\end{figure}

\begin{figure}[h]
\centering
\begin{subfigure}[b]{.49\columnwidth}
  \centering
  \includegraphics[width=\textwidth]{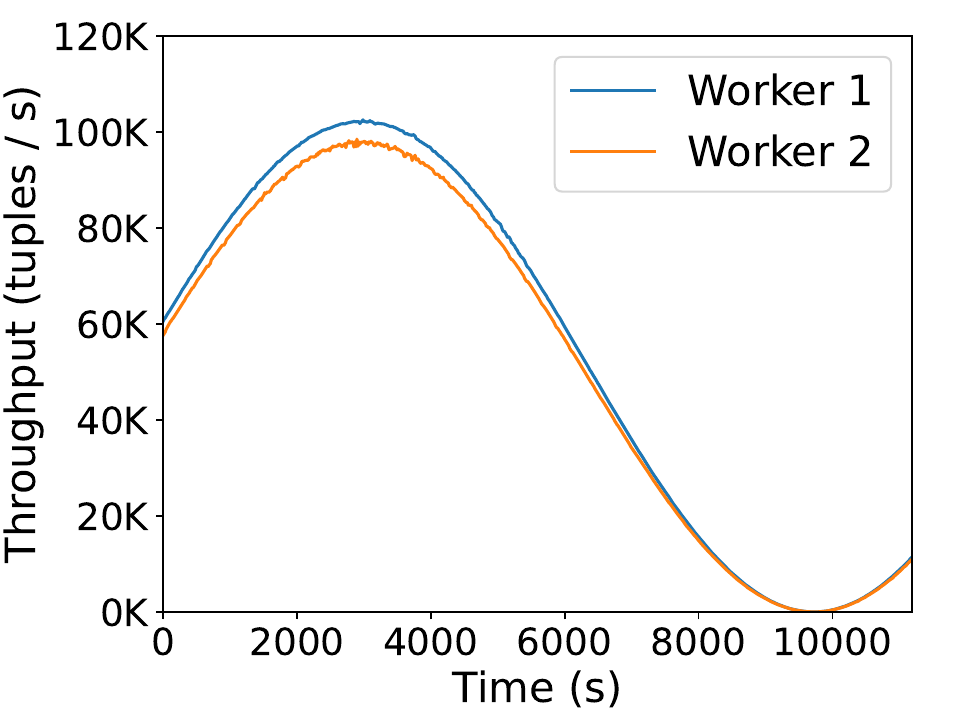}
  \caption{Throughput per worker}
  \label{fig:sine-throughput}
\end{subfigure}
\begin{subfigure}[b]{.49\columnwidth}
  \centering
  \includegraphics[width=\textwidth]{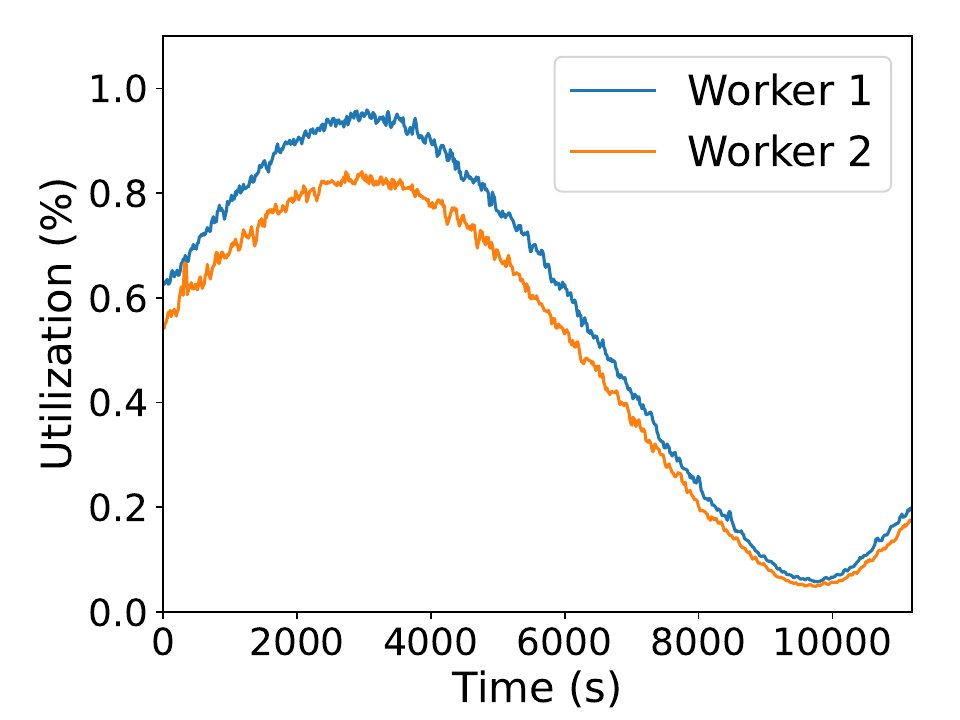}
  \caption{CPU utilization per worker}
  \label{fig:sine-cpu}
\end{subfigure}
\caption{Proportional data skew over CPU utilization}
\label{fig:sine-p2}
\end{figure}

\autoref{fig:data_skew} shows metrics from a DSP job with stateful operators processing tuples at maximum capacity with a parallelism of 12. In this example, the data is generated randomly across 100 keys, and each worker reads from its own Kafka partition. In theory, the keys could be almost evenly distributed among parallel operators, with each worker handling eight or nine keys. However, as can be seen in~\autoref{fig:data_skew_a} and~\autoref{fig:data_skew_b}, data skew is apparent and the workers display a spectrum of throughput and CPU utilization. Although a worker using only 75\% CPU is theoretically capable of processing more tuples, it cannot receive more tuples due to how the keys are distributed. Its maximum capacity is thus capped at its throughput at 75\% CPU utilization. 
As seen in~\autoref{fig:sine-p2}, data skew across workers remains proportional at different levels of throughput and is most prominent at high CPU utilization.
Following these observations, in general, the maximum capacity of a worker is limited by its proportion to the worker with the highest CPU utilization.

\begin{figure}
\centering
\begin{subfigure}{.49\columnwidth}
  \centering
  \includegraphics[width=\textwidth]{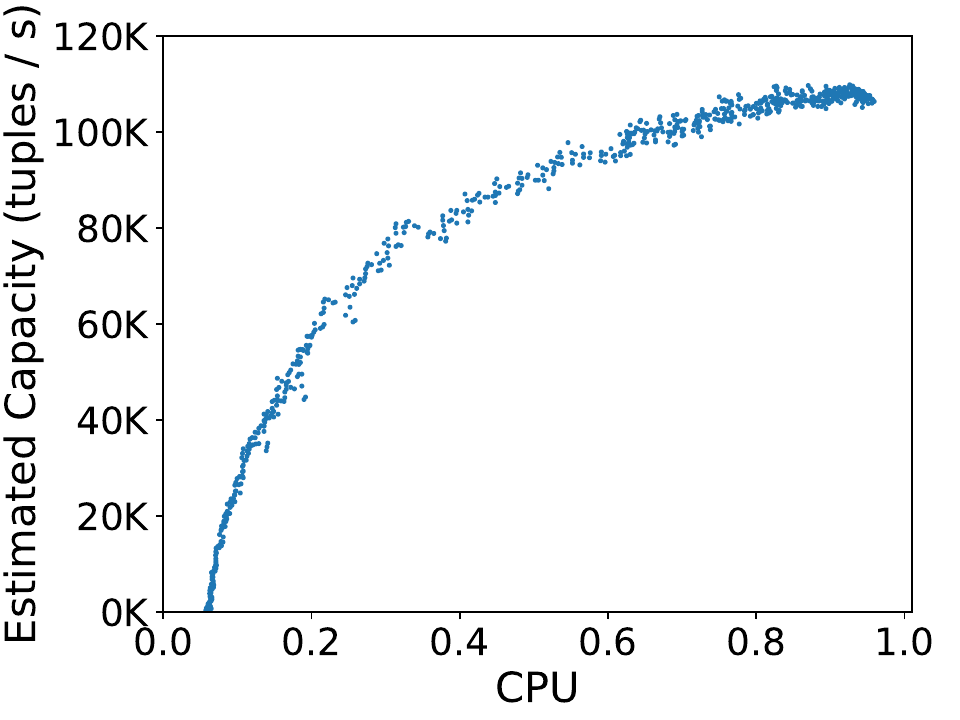}
  \caption{Estimated capacity}
  \label{fig:simple-capacity}
\end{subfigure}%
\begin{subfigure}{.49\columnwidth}
  \centering
  \includegraphics[width=\textwidth]{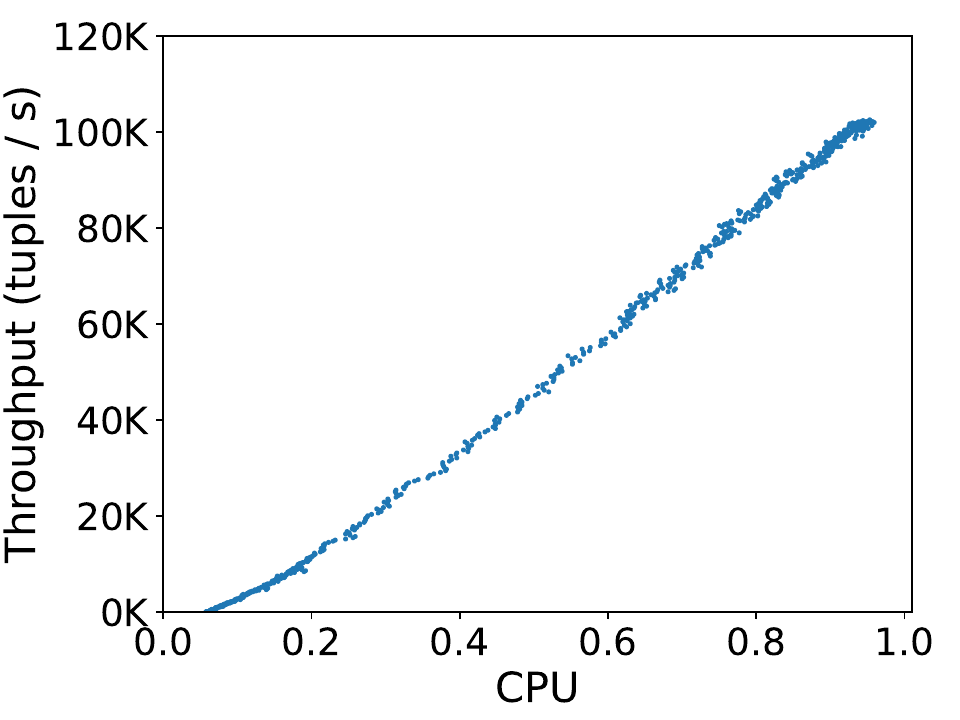}
  \caption{Observed throughput}
  \label{fig:linear-regression}
\end{subfigure}
\caption{Capacity over CPU utilization}
\label{fig:estimating-capacity}
\end{figure}

A worker's theoretical maximum capacity corresponds to how many tuples it can process at 100\% CPU utilization. 
The calculation
\begin{equation*}
\text{Capacity} = \frac{\text{Throughput}}{\text{CPU Utilization}}
\end{equation*}
yields a quick estimation of a worker’s maximum capacity. However, as can be seen in~\autoref{fig:simple-capacity}, the accuracy of this estimation is highly dependent on the level of CPU utilization.
Although this simple capacity calculation provides a reasonable estimation when CPU utilization is greater than 70\%, and the maximum capacity estimation at a given CPU utilization is higher than the observed throughput, a more accurate capacity estimation is needed to inform stable scaling decisions.
Given the linear relationship between throughput and CPU utilization, seen in~\autoref{fig:linear-regression}, linear regression lends itself well to this task.
Because only one explanatory variable is required to predict a worker's maximum throughput, it is possible to build a simple regression model 
using an efficient online analytical calculation.
Using throughput as the dependent variable $y$ and CPU utilization as the independent variable $x$, the throughput at a given CPU utilization can be predicted using the simple linear regression formula $y=\alpha + \beta x$, where $\alpha$ is the y-intercept and $\beta$ is the slope.
The slope can be calculated by dividing the covariance of CPU ($X$) and throughput ($Y$) observations by the variance of CPU observations. The y-intercept is given by subtracting the slope multiplied by the mean of all CPU observations ($\overline{X}$) from the mean of all throughput observations ($\overline{Y}$). 
Put together, this yields the following equation to predict the capacity at a desired CPU utilization $\text{CPU}_{\text{desired}}$:

\begin{equation*}
\text{Capacity} = \overline{Y} - \frac{\text{cov}(X, Y)}{\text{var}(X)} \cdot \overline{X} + \frac{\text{cov}(X, Y)}{\text{var}(X)} \cdot \text{CPU}_{\text{desired}}
\end{equation*}

To update such a model with new observations, the running covariance, variance, and means can be computed using an adaptation of Welford's online algorithm for calculating variance~\cite{Welford1962NoteOA}. The algorithm passes over new observations once, updating the count of observations and the delta for the new mean CPU and throughput observations. These values are then used to update the variance, covariance, CPU, and throughput means. Welford's algorithm is numerically stable and all required values can be computed on one pass of the data, meaning that there is no need to save observations, which could require much storage space over the course of time. %

A linear regression model is computed for each worker individually to increase the accuracy of the overall capacity estimation at a given scale-out. To find the capacity for each worker while accounting for data skew, the linear regression model can predict the value at the expected maximum CPU utilization. As described previously, the expected maximum CPU utilization of a worker is proportional to the worker with the maximum CPU. 

Daedalus differentiates capacity estimations at seen and unseen scale-outs. 
The capacity at the current scale-out is calculated by summing the estimated capacity across all workers. The estimation for a current scale-out can accurately assess how data is distributed among workers. The capacity at other scale-outs is estimated using the average capacity multiplied by the scale-out. While this can not guarantee how data will be distributed at that scale-out, it provides an adequate heuristic. When possible, Daedalus uses previously observed capacity estimations over purely predicted estimates for seen scale-outs.
Ideally, the regression models would have a range of CPU observations in order to be more robust and accurate. However, due to the low variance present in the CPU-throughput regression (as seen in~\autoref{fig:linear-regression}), relatively few data points are needed to accurately estimate capacity. From experimentation, the regression model is able to accurately estimate capacity in as little as 60 seconds, the time of a single loop. This observation holds across different jobs and various scale-outs.

\subsection{Scaling Decisions}

Central to any autoscaling approach is how it makes scaling decisions. 
While reactive approaches benefit from being able to use real observations to inform scaling decisions, the point at which this data is available can lead to QoS violations until an acceptable scale-out is deployed. On the other hand, proactive approaches need to deal with the uncertainty of predicting the future. Daedalus uses a hybrid of reactive and proactive approaches, using both observed data and future forecasts in order to reactively scale in and proactively scale out. 

In a reactive manner, Daedalus uses historical workload data since the last iteration of the MAPE-K loop to find the minimum scale-out needed to process the workload. This helps to offset uncertainty from inaccurate future predictions, such as when future forecasts are lower than the actual workload. Using TSF the workload can be anticipated to enable proactive scaling decisions. TSF grants the ability to scale out before capacity is exceeded, helping to reduce QoS violations and minimize end-to-end latency. It also enables making long-lived scaling decisions, as a scale-out can be chosen that can handle the current and future workload. It also allows to more accurately calculate recovery time, instead of needing to assume that the workload will remain constant.

In each iteration of the MAPE-K loop, the workload since the last iteration is collected, capacities are estimated for all scale-outs, and the future workload is predicted. With this information, it can be determined if rescaling is necessary and if so, how to rescale.

The pseudocode for determining the appropriate scale-out is shown in~\autoref{alg:scaleout}. 
Because Daedalus aims to make long-lived scaling decisions, the algorithm first checks if rescaling is absolutely necessary in case a rescale recently occurred. If a rescale was done in the last ten minutes, it checks that the current capacity can handle the average observed workload and maximum future workload until the next loop iteration. The average workload is used instead of the maximum in order to remove noise from the actual workload, such as any spikes that may have occurred. In case both of these conditions are true, the algorithm returns and no rescaling is necessary.

    \begin{algorithm}
    \DontPrintSemicolon
    \caption{An algorithm to determine the scale-out}
    \label{alg:scaleout}
    
    \SetKwComment{Comment}{\# }{ }
    \KwData{C := Capacities, W := Workload, TSF}
    \SetKwFunction{predictRecoveryTime}{predict\_recovery\_time}

    \If{time since last rescale $<$ \text{600s}}{
        \If{C$_{current}$  $>$ W$_{avg}$ and TSF$_{max}$ until next loop}{
            \Return{current parallelism}
        }
    }

    \For{$i = 1 $ \KwTo MaxScaleout}{
      \If{C$_i$ $>$ W$_{avg}$}{
        $RT_i \gets $ \predictRecoveryTime{i}\;
        \If{$RT_i > RT_{target}$}{
            \textbf{continue} %
        }
        \If{$C_i < $ TSF$_{max}$ until $RT_i$}{
            \textbf{continue} %
        }

        \If{i = current parallelism}{
            \Return{i} 
        }

        \If{i $<$ current parallelism and $C_i$ $<$ consumer lag}{
            \textbf{continue}
        }

        \If{C$_i$ $>$ TSF$_{max}$}{
            \Return{i}
        }
    
      }
    }
    \Return{MaxScaleout} %
    
    \end{algorithm}

The algorithm then iterates over all possible scale-outs to find the lowest number of workers that can process the incoming workload for the next 15 minutes and ensures that recovery is possible within a target time. In the first step, it is checked that the scale-out is capable of processing the average observed workload, similar to reactive autoscaling approaches. This prevents assessing the validity of smaller scale-outs that cannot handle the incoming workload, which would produce an inaccurate scaling decision.

Next, the recovery time is predicted for the scale-out. The recovery time estimates the time needed to process accumulated tuples while the system is down until it can catch up to a normal state. This handles cases for scaling in and out (when the examined scale-out is different from the current scale-out) as well as failure (when the scale-out is the same as the current scale-out). If the estimated recovery time is greater than the specified target recovery time, the currently investigated scale-out is invalid and the next scale-out is examined. It is also checked that the scale-out can handle the future workload while recovering. Otherwise, a rescale would be necessary while the system is recovering. If this is the case, the next scale-out is examined.

At this point in the algorithm, the examined scale-out is valid. In case the current parallelism is the same as the examined scale-out, the algorithm returns and no rescaling is necessary. Otherwise, the examined scale-out requires scaling in or out.
To prevent scaling in too early, the consumer lag is investigated. In case the consumer lag is larger than the examined capacity, it is a good indication that the system is recovering or potentially overloaded. Even though the target recovery time would be met when scaling in, it is beneficial to wait until the system has caught up to provide better end-to-end latencies. Therefore, the next scale-out is examined, potentially pushing the decision to scale in until the next iteration.

Lastly, to ensure that the scale-out will be long-lived, it is checked that the capacity is greater than the maximum of the full TSF prediction of 15 minutes. Doing so also helps to minimize the need to scale again after 10 minutes, making the first check of the algorithm a precautionary measure.
After a scaling action has been initiated, the system is given three minutes to adjust to the new scale-out before another scaling action can occur. This allows the system to stabilize and helps prevent flapping, where autoscaling alternates between scaling in and out. Including a grace period is a common practice. For example, Kubernetes Horizontal Pod Autoscaler (HPA) uses a default stabilization time of five minutes. Since Daedalus anticipates the future workload and scaling decisions are designed to be long-lived, this arbitrary threshold should have little impact. However, because no scaling actions can be taken in this period, it increases the need to generate accurate scaling decisions.

\subsection{Time Series Forecasting}\label{section:tsf}
TSF is used to anticipate future workloads and enables proactive scaling decisions that help to make long-lived scaling decisions, reduce the overhead cost of rescaling, and minimize latencies. Despite DSP workloads being dynamic and therefore difficult to predict, TSF has been shown to improve autoscaling decisions to reduce resource consumption and better meet QoS requirements~\cite{Gontarska2021EvaluationOL,Geldenhuys2022PhoebeQD}. 
Multiple TSF methods exist including ARIMA, Holt-Winters, FB Prophet, and deep learning methods. TSF models often need to be configured themselves, which can be challenging and require expert knowledge. %
Fortunately, models such as auto-ARIMA exist that automatically find optimal parameters based on provided workload data. Using the pmdarima\footnote{\url{https://alkaline-ml.com/pmdarima/index.html}\LastTimeAccessed} library, an initial model is trained with the available workload, and the model is updated with the latest workload observations in every iteration of the MAPE-K loop. A new forecast is generated in each loop to predict the next 15 minutes of the workload at second-level granularity.

Though ARIMA has been extensively evaluated and has been found to produce good forecasts for up to 15 minutes~\cite{Gontarska2021EvaluationOL}, it is essential to evaluate the quality of the forecasts and include a mechanism to retrain models that consistently deliver poor predictions.
In each iteration of the MAPE-K loop, the latest workload metrics are collected and a new forecast is generated. At this point, it is possible to compare the previous forecast with the latest workload observations. To gain an insight into the overall accuracy of the forecast, the weighted absolute percentage error (WAPE) is used, which is calculated by weighting the error between the actual values and forecasts with the sum of the actual values over all units of time.
A lower WAPE indicates a better forecast. In case the previous forecast was inaccurate with respect to the workload, it is possible that the next forecast will be similarly inaccurate. If this happens, the ARIMA forecast is replaced with a forecast that applies a simple regression on the workload. This heuristic uses the slope from the latest workload observations and projects the workload 15 minutes into the future. This linear approximation only provides a fallback forecast when the previous TSF prediction was poor. If the TSF predictions are consistently poor for 15 consecutive iterations, the TSF model is retrained in a background thread to not interrupt the execution of the MAPE-K loop. Once training is completed, the newly trained model replaces the existing one.

\subsection{Recovery Time}\label{section:recoverytime}

\begin{figure}[h]
    \centering
   \includegraphics[width=\linewidth]{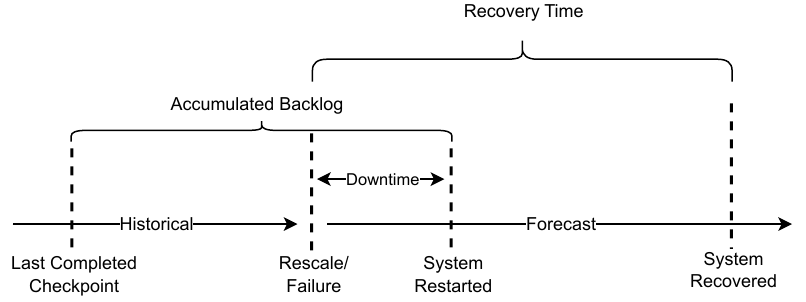}
   \caption{Predicting recovery time}
   \label{fig:recovery-time-prediction}
\end{figure}

Recovery time is a vital, yet often overlooked, aspect of autoscaling DSP systems. DSP systems must be fault tolerant and continue to operate despite failures that are likely to occur over the course of a long-running job. In addition to failures, when rescaling, processing must often be stopped temporarily in order to start new workers and recompute data parallelism among stateful operators. To ensure exactly-once processing, the DSP system must reprocess tuples that occurred after the last completed checkpoint as well as process tuples that arrived while the system was down. In addition to this accumulated backlog, tuples will continue to arrive while the system is catching up. As depicted in~\autoref{fig:recovery-time-prediction}, recovery time describes the time from when processing stops, due to rescaling or failure, until the system has caught up and processed the accumulated backlog. Only then can normal processing resume.

Recovery time has a direct impact on end-to-end latency. While the DSP system is down tuples cannot be processed, increasing their end-to-end latency. While the system is recovering, the accumulated backlog must be processed before the tuples that continue to arrive can be processed, creating a cascading effect. Therefore, by incorporating a target recovery time into autoscaling decisions, latency can be better minimized by avoiding long recovery times.

Recovery time can be predicted by calculating the accumulated backlog and estimating how long it will take to process the backlog in addition to the incoming workload using the extra processing capacity of the targeted scale-out.

The accumulated backlog of tuples is given by the tuples that need to be reprocessed since the last completed checkpoint and tuples that arrive while the system is down. For calculating the number of tuples since the last checkpoint, the worst case is assumed, for example, 10 seconds for a 10 second checkpoint interval. The number of tuples to be reprocessed can therefore be calculated by taking the number of tuples that occurred in the last checkpoint interval seconds using the historical workload. The worst case is assumed in order to provide a comparative baseline regardless of when the last checkpoint actually occurred with respect to the prediction. Failures and rescaling can occur at any time. Assuming the worst case also results in a larger recovery time prediction than would be expected on average, which provides a buffer to better achieve the target recovery time.

To estimate the number of tuples that arrive while the system is down, it is necessary to anticipate the time the system is down and use the workload forecast. The anticipated downtime is initially set to 30 seconds for scaling out and 15 seconds for scaling in. However, this value can be adaptively updated by monitoring the actual recovery time, discussed in the next section. This generally yields more accurate recovery time predictions over time. 

After the DSP system restarts, it can begin processing the accumulated backlog using the maximum capacity of the target scale-out. While doing so, tuples will continue to arrive. Though the system will process the backlog first, when determining the point at which the system is caught up, the order tuples are processed is irrelevant. The end of the recovery time can thus be determined using the extra capacity, forecast, and accumulated backlog. The extra capacity available at the target scale-out can be obtained by subtracting the forecast from the capacity. For each future step, it is then checked if the cumulative extra capacity exceeds the accumulated backlog. When this is the case, the system has recovered, and the predicted recovery time can be returned.

\subsection{Monitoring with Anomaly Detection}

To improve the accuracy of recovery time predictions, the actual recovery time after a scaling action is observed using statistical anomaly detection. When monitoring recovery time, the goal is to identify when processing returns back to normal after rescaling. Anomaly detection is therefore a fitting paradigm, as it provides a means for classifying normal and abnormal behavior. 
Because the objective is to find points when the system's throughput deviates from the incoming workload, it is sufficient to use statistical anomaly detection on the difference between the workload and throughput. 
Statistical anomaly detection classifies observations as anomalous if the distance of an observation to the mean is above a certain threshold.
Daedalus uses the threshold of one standard deviation. The anomaly detection model keeps track of the job's running mean and variance of the difference between workload and throughput using Welford's previously mentioned online algorithm. 
After a rescaling action, Daedalus checks for anomalies until the system has recovered. Because it takes time for a system to recover, the anomaly detection monitoring is run in a background thread to prevent interfering with the MAPE-K loop.

\subsection{Implementation}

Daedalus uses the proven MAPE-K control loop for self-adaptive systems to provide structured execution. The monitor, analyze, plan, and execute phases are structured as follows:

\begin{itemize}
    \item \textbf{Monitor}: Daedalus collects metrics from Prometheus. From the DSP system, it collects the throughput for each worker measured by the number of records consumed by the source operator, CPU utilization of each worker using a moving average of one minute to reduce noise, and the overall consumer lag representing available, but not processed, tuples. For convenience, the job up-time and current parallelism are also collected. From the data source, the incoming workload rate is collected and measured in tuples per second.

    \item \textbf{Analyze}: The maximum capacity of each worker is calculated using CPU-throughput regression models, and the capacity for each scale-out is estimated. The ARIMA TSF and anomaly detection models are updated, and the future workload is predicted using the TSF model. 

    \item \textbf{Plan}: The optimal scale-out is determined using the algorithm described in~\autoref{alg:scaleout}. The chosen scale-out must be able to process the incoming workload and recover within the target recovery time. If rescaling is necessary, the scale-out must also be able to process the predicted workload. 
    
    \item \textbf{Execute}: Any planned scaling action is executed by the Kubernetes client. The actual recovery time is then monitored with anomaly detection in a background thread.
    
    \item \textbf{Knowledge}: Knowledge represents the shared information between models. This is the collected metrics, capacity models, forecasts, anomaly detection, scaling actions, and recovery time information.
\end{itemize}

The MAPE-K loop runs every 60 seconds and takes on average one second to execute because of its low computational complexity.

\section{Evaluation}
\label{sec:evaluation}

To demonstrate the effectiveness of Daedalus, it is evaluated with three DSP jobs and two DSP systems, Flink and Kafka Streams. The next sections describe the DSP jobs, the comparison systems, experimental setup, and results. This chapter concludes with a discussion that evaluates the overall performance of Daedalus.

\subsection{DSP Jobs}

Daedalus is evaluated with three representative DSP jobs: WordCount, Yahoo Streaming Benchmark, and Traffic Monitoring.
All relevant code can be found in the Daedalus GitHub repository\footnote{\url{https://github.com/dos-group/daedalus}}.

\subsubsection{WordCount}

WordCount is a popular DSP job that is readily available for different DSP systems, as it frequently serves as the exemplary tutorial to illustrate stream processing. Because of its simplicity and ubiquity, WordCount is often used to compare DSP systems~\cite{Bordin2020DSPBenchAS}.
WordCount computes a running total of word occurrences in a given text corpus. It takes lines of text as input, splits the line into words, and returns each word along with its cumulative word count. In order to test the job with dynamic workloads, WordCount has been modified to read input from a Kafka source. The output of words and their count are written to a console sink.

\subsubsection{Yahoo Streaming Benchmark}

The Yahoo Streaming Benchmark was one of the first benchmarks to evaluate major modern DSP systems including Apache Storm, Apache Spark, and Apache Flink~\cite{Chintapalli2016BenchmarkingSC}.
Though created in 2016, it continues to be used to compare DSP systems and the pipeline serves as a baseline for further benchmarking jobs~\cite{OpenBenchmark2020}. The pipeline features representative operations of stream processing jobs such as filtering, windowing, aggregation, and joining data with a database.
The job is an advertising analytics use case that consists of deserializing JSON ad events from a Kafka source, filtering ads based on an event type and removing unnecessary fields, matching the ad to a campaign ID stored in Redis, and counting the number of times an ad was viewed within a ten second tumbling window. In the original benchmark, read and write operations to Redis became a bottleneck when operating at larger scales.
For this reason, the job has been modified so that campaign IDs from Redis are cached in the DSP job and the resulting ad counts are written to a Kafka sink instead of Redis.

\subsubsection{Traffic Monitoring}

The Traffic Monitoring job is an IoT use case that calculates the average speed of moving vehicles in a particular radius adapted from the IoT Vehicles Experiment~\cite{Geldenhuys2021KhaosDO}. 
The job reads JSON vehicle events from a Kafka source, filters out events not contained within a radius of interest, calculates the average speed of vehicles in a ten second tumbling window, and enriches the vehicle information before outputting to a Kafka sink.

\subsection{Workload Generation}

In order to test the effect of dynamic workloads on autoscaling approaches, each job is run with a generator that produces a configurable amount of tuples per second to a Kafka topic. The data generators use the Akka
actor system to simulate tuples in a highly scalable way. Generators are run inside the Kubernetes cluster to increase scalability and reduce the impact of network latency.

Each job is tested with a representative workload. %
Given its artificial nature, the workload for the WordCount job is a sine wave with two periods. The Yahoo Streaming Benchmark workload is taken from realistic online advertising click-through rate data\footnote{\url{https://www.kaggle.com/competitions/avazu-ctr-prediction}\LastTimeAccessed}. 
Lastly, the traffic monitoring workload was generated based on the TAPASCologne scenario and SUMO to simulate realistic traffic patterns in the city of Berlin~\cite{Geldenhuys2021KhaosDO}.
Each job was benchmarked to determine the maximum throughput achievable with 12 workers.
All workloads have been scaled so that the maximum number of tuples is less than this throughput in order to more fairly compare autoscaling approaches to a static scale-out with 12 workers.
Additionally, workloads are scaled to a duration of 6 hours. With these parameters, the workloads allow for a range of scaling decisions to test autoscaling approaches.

\subsection{Comparison Systems}
In order to demonstrate its usefulness, Daedalus is compared to a static deployment capable of processing the peak workload, HPA native to Kubernetes, and Phoebe, a recent state-of-the-art approach conceptually comparable to Daedalus.

\subsubsection{Static Deployment}
To serve as a baseline, a static deployment with 12 workers is deployed. As previously stated, this scale-out is capable of processing the peak workload for each job. It can therefore indicate if autoscaling approaches over-provision resources. While it is likely to have the highest resource consumption, it should also provide stable latencies because it will not rescale. It therefore provides a baseline latency comparison as well as showing the potential reduction of resource usage achieved through autoscaling.

\subsubsection{Horizontal Pod Autoscaler}
Kubernetes provides a built-in method for automatically scaling resources called a Horizontal Pod Autoscaler (HPA). A HPA monitors one or more metrics, such as CPU utilization or memory, and horizontally scales the target deployment in accordance with a user-defined policy. %
By default, the HPA monitors whether metrics violate the defined thresholds every 15 seconds. The HPA ignores instances that have not started yet in its calculation and waits for a default of five minutes between performing scaling actions to avoid flapping. 
While HPAs are commonly used and intuitive to understand, choosing a reasonable threshold to fairly compare with Daedalus is not simple. In line with Daedalus's objectives, the DSP system should target a high utilization to process the incoming workload with minimal resources while leaving extra processing capacity to recover from failure or rescaling actions. As shown in~\autoref{fig:data_skew}, a system operating at full capacity does not necessarily use 100\% CPU. Though Daedalus does not use CPU utilization thresholds to trigger scaling decisions, most scaling decisions occurred between 80\% and 85\% when testing with Flink. Therefore, two HPA deployments are tested. One targeting 80\% utilization, and one targeting 85\% utilization. With Kafka Streams, the HPA deployments target 60\% and 80\% utilization.

\subsubsection{Phoebe}
Phoebe is an approach fairly similar to Daedalus, and unlike many other DSP autoscaling approaches, its source code is publicly available\footnote{\url{https://github.com/dos-group/phoebe}\LastTimeAccessed}.
Phoebe uses initial profiling runs to build QoS models and TSF to better meet QoS requirements. Similarly to Daedalus, it is capable of choosing worker parallelism in accordance with a target recovery time. 
Unlike Daedalus though, Phoebe explicitly models and accounts for latency.
It also injects failures into profiling runs, measures the resulting recovery times, and incorporates them into its QoS models.
Phoebe was implemented using Flink and also evaluated using the Yahoo Streaming Benchmark. 

\subsection{Experimental Setup}

Experiments were run on a five-node Kubernetes
cluster with ample CPU and memory, totaling 160 cores and 640 GB RAM. 
Details can be seen in~\autoref{tbl:clusterspecs}. 
The experiments make use of Apache Kafka
as a data source and sink. Kafka topics have been created with the same number of partitions as the maximum scale-out so that each worker can consume from its own partition. Prometheus
is used as a time series database and periodically scrapes metrics from Kafka and the target DSP system. Redis
is used as a data store for the Yahoo Streaming Benchmark. Lastly, HDFS
is used by Flink for saving checkpoints to storage. 

All approaches are deployed at the same time and read from the same Kafka source topic. 
For Flink deployments, each approach is deployed in application mode, ensuring resource isolation as there is no shared JobManager. The Flink deployments also make use of Flink's recently introduced reactive-mode,\footnote{\url{https://flink.apache.org/2021/05/06/reactive-mode.html}\LastTimeAccessed} a built-in method that allows elastic scaling. With reactive mode, Flink will automatically scale to the number of desired replicas and rescale the job from the last completed checkpoint. Defining the desired number of replicas is done using Daedalus or HPA. For both Flink and Kafka jobs, a custom metric was added to measure end-to-end latency. This metric measures the time from tuple generation until the end of processing, ignoring windowing periods. The 95th percentile latency is reported in the results.
Across all our setups, we provided workers with significant memory access to preemptively ensure that memory constraints did not become a bottleneck. 
This was validated through empirical observations.

Approaches are evaluated by the quality of their scaling decisions. All jobs use exactly-once processing semantics and the DSP systems process all tuples. Therefore, end-to-end latency is a more suitable metric to indicate the processing performance. End-to-end latencies are reported by an empirical cumulative distribution function. 
Average latencies are also reported, though they can be misleading, since peaks of high latency during rescaling can skew an average upwards. While some papers exclude rescaling downtime in result calculations, this is misleading as recovery time is a crucial factor. Therefore, no metrics are excluded in the results.

\begin{table}
\centering
    \begin{tabular}[t]{rp{0.65\linewidth}}
        \toprule
        Resource&Details\\
        \midrule
        OS       & Ubuntu 20.04.3\\
        CPU      & AMD64 Processor, 32 cores, 2.8GHz\\
        Memory   & 128 GB RAM\\
        Storage  & 3TB RAID0 (3x1TB, Linux software RAID)\\
        Network  & 10 GBit Ethernet NIC\\
        Software & Kubernetes v1.24.3, Docker v19.3, Java v1.11, Flink v1.16.0, Kafka v3.2, ZooKeeper v3.8, HDFS v2.8, Redis v6.2.7, Prometheus v2.39.1
        \\
        \bottomrule
    \end{tabular}
\caption{Cluster specifications}
\label{tbl:clusterspecs}
\end{table}

\subsection{Flink Experiment Results}

This section contains the results for the Wordcount, Yahoo Streaming Benchmark, and Traffic Monitoring experiments run using Flink as the target DSP system. For these experiments, Daedalus used a recovery time of 600 seconds. 
Each experiment was executed five times to ensure consistency of the results.
As the form of presentation of the results is repeated, we briefly explain the individual subplots on the basis of~\autoref{fig:wc_results}: \autoref{fig:wc_results_workload} shows the workload over time; \autoref{fig:wc_results_parallelism} illustrates the number of workers over time as a result of the respective scaling method; \autoref{fig:wc_results_latencies} presents the cumulative probability of observed latencies, which allows for statements such as \emph{"more than 80\% of all latency values where smaller or equal $10^3$ms for Daedalus"}; Lastly, the total resource usage, normalized with respect to the static baseline, is displayed in~\autoref{fig:wc_results_resources}.

\subsubsection{WordCount Results}

\begin{figure}
\centering
\begin{subfigure}[b]{.49\columnwidth}
  \centering
  \includegraphics[width=\textwidth]{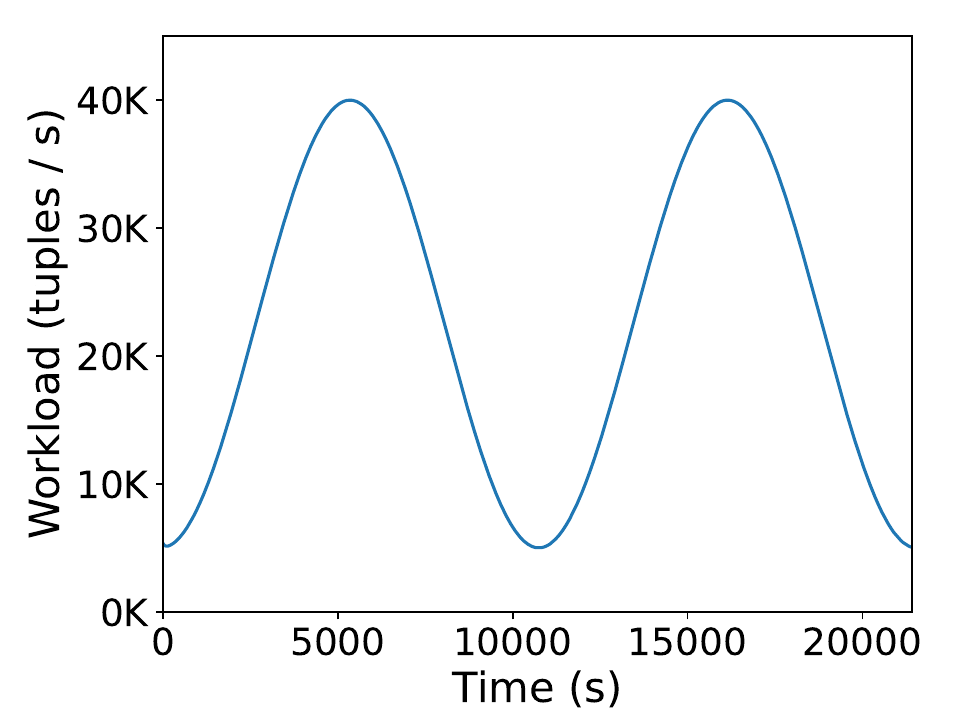}
  \caption{Workload}
  \label{fig:wc_results_workload}
\end{subfigure}
\begin{subfigure}[b]{.49\columnwidth}
  \centering
  \includegraphics[width=\textwidth]{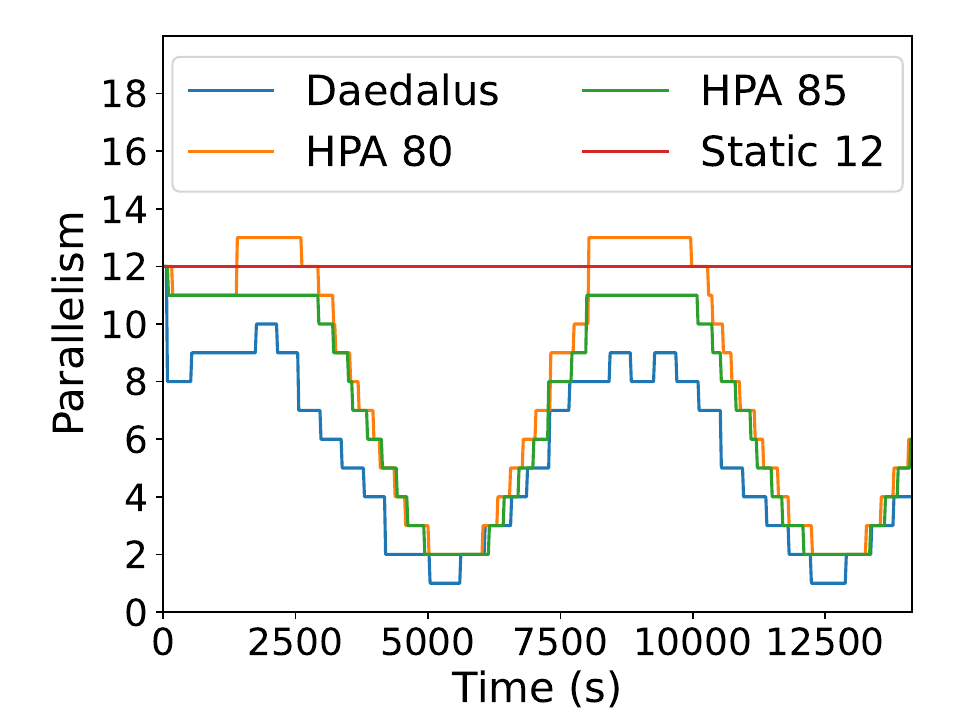}
  \caption{Number of workers}
  \label{fig:wc_results_parallelism}
\end{subfigure}
\begin{subfigure}[b]{.49\columnwidth}
  \centering
  \includegraphics[width=\textwidth]{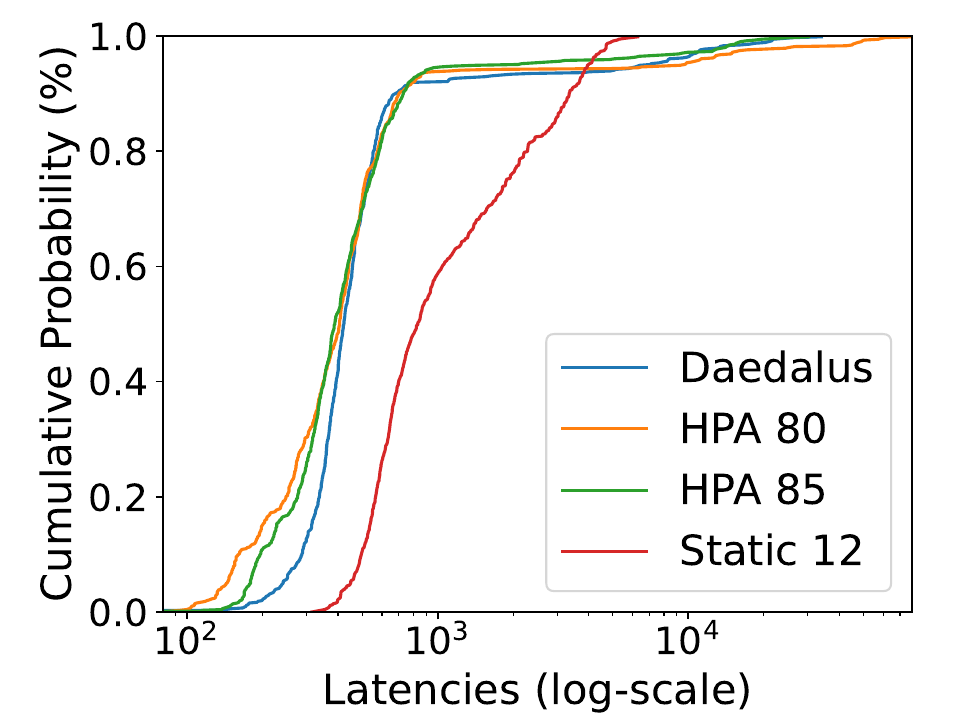}
  \caption{Latency distribution}
  \label{fig:wc_results_latencies}
\end{subfigure}
\begin{subfigure}[b]{.49\columnwidth}
  \centering
  \includegraphics[width=\textwidth]{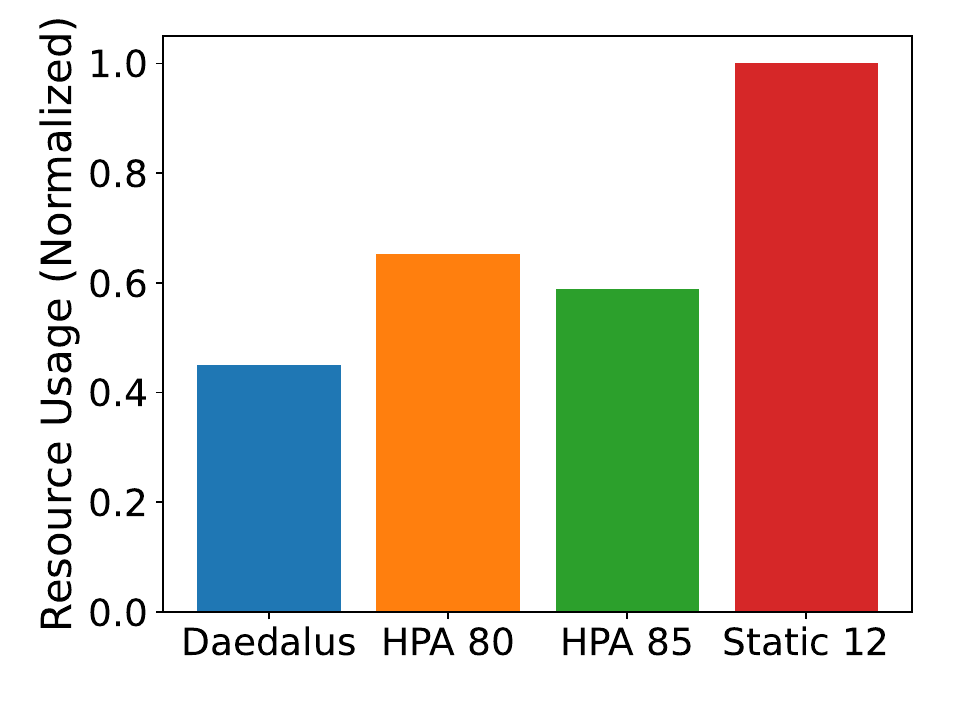}
  \caption{Resource usage}
  \label{fig:wc_results_resources}
\end{subfigure}
\caption[WordCount results]{Flink WordCount results}
\label{fig:wc_results}
\end{figure}

The results for the WordCount job are shown in~\autoref{fig:wc_results}. Despite the job's simplicity, WordCount is highly susceptible to data skew, making it challenging to accurately estimate capacity for unseen workloads. In practice, this also means that the maximum observed capacity of workers at a specific scale-out can vary after rescaling to that scale-out again later in the job. Nevertheless, all autoscaling approaches were able to match the resources to the workload and process tuples in a timely manner. 
In general, the autoscaling approaches scale out around the same time, but Daedalus is able to scale in faster than the HPA methods.

As can be seen in~\autoref{fig:wc_results_latencies}, generally, all approaches perform very similarly with most latency measurements falling between $10^2$ms and $10^3$ms. Average latencies over the span of the job are as follows: Daedalus with 1,171 ms, HPA 80 with 1,791 ms, HPA 85 with 961 ms, and Static 12 with 1,408 ms. Notably, the static scale-out has proportionally slightly higher latencies. As found in previous research~\cite{Zhang2021AuTraScaleAA} and shown in~\autoref{fig:metric_relationships_d}, over-provisioning resources does not guarantee optimal latencies. Proportionally, the autoscaling approach latencies are quite similar. The larger increase in latencies above the 95\% mark indicates when the systems are temporarily unavailable due to autoscaling. 

When comparing resource utilization, Daedalus uses significantly fewer resources. On average, Daedalus used 5.4 workers, HPA 80 used 7.8, HPA 85 used 7.0, and the static scale-out naturally used 12. Overall, Daedalus used 55\% less resources than the static scale-out, 31\% less resources than the HPA 80, and 23\% less resources than the HPA 85. So with fewer resources, Daedalus was able to achieve comparable latencies to HPA. 

\subsubsection{Yahoo Streaming Benchmark Results}

\begin{figure}
\centering
\begin{subfigure}[b]{.49\columnwidth}
  \centering
  \includegraphics[width=\textwidth]{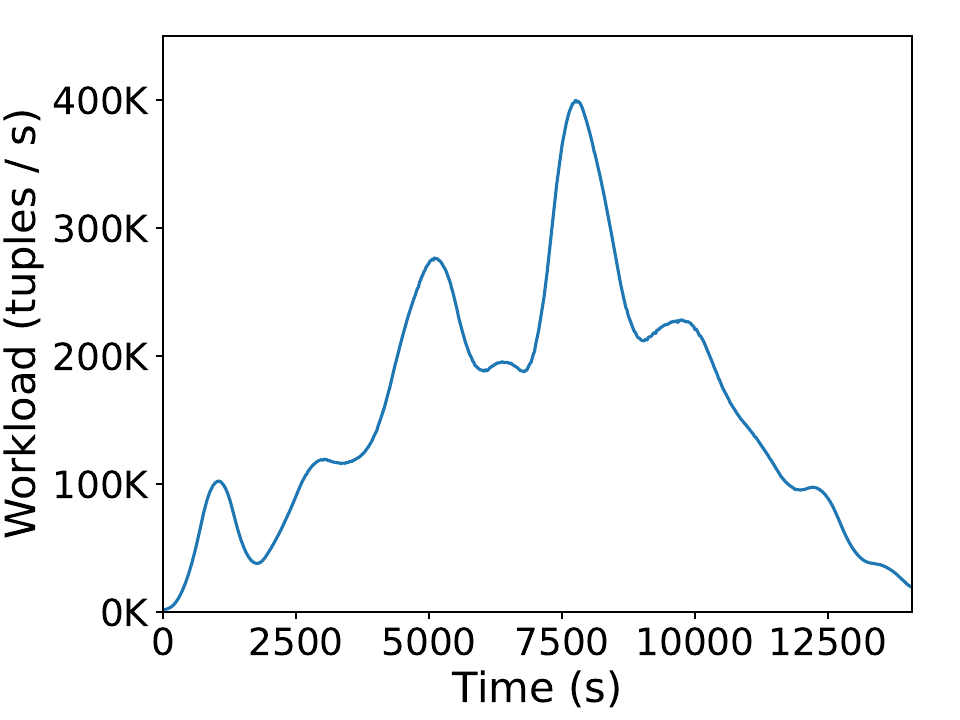}
  \caption{Workload}
  \label{fig:ysb_results_workload}
\end{subfigure}
\begin{subfigure}[b]{.49\columnwidth}
  \centering
  \includegraphics[width=\textwidth]{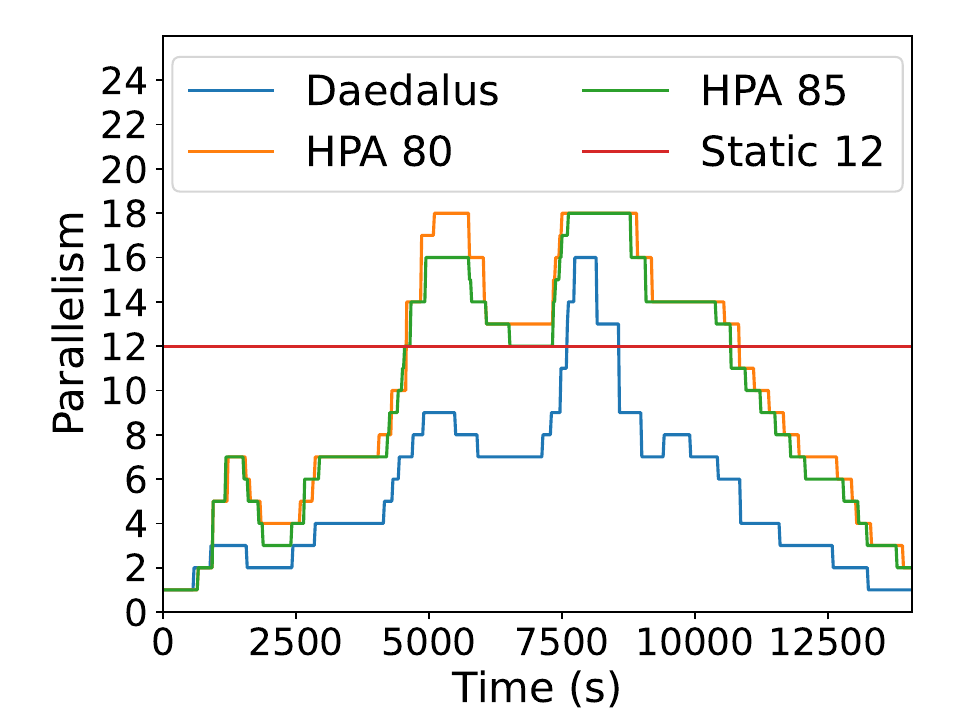}
  \caption{Number of workers}
  \label{fig:ysb_results_parallelism}
\end{subfigure}
\begin{subfigure}[b]{.49\columnwidth}
  \centering
  \includegraphics[width=\textwidth]{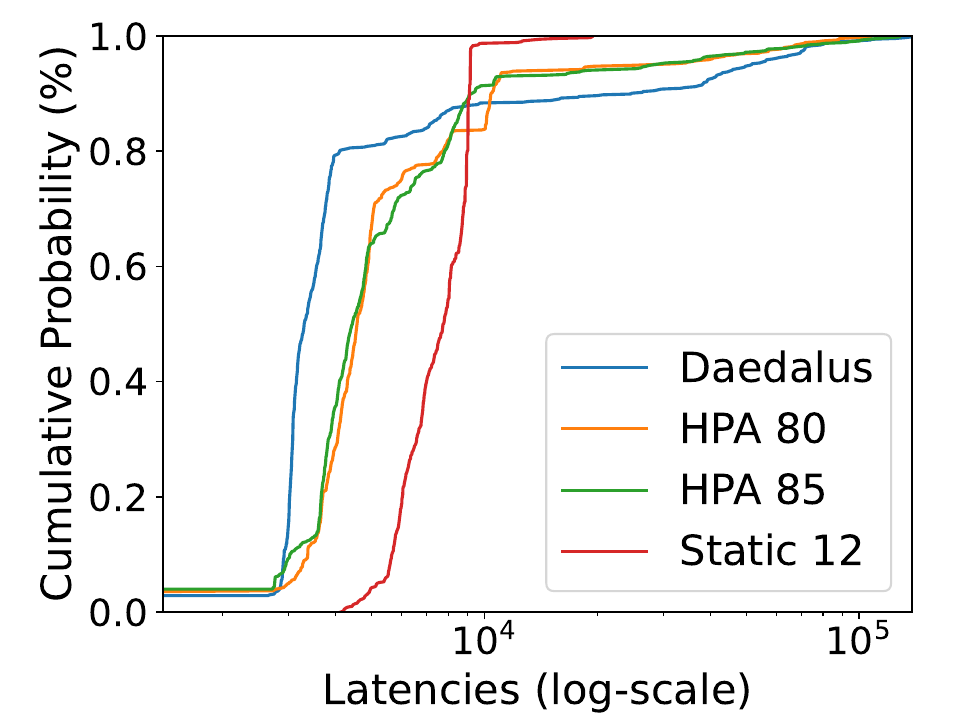}
  \caption{Latency distribution}
  \label{fig:ysb_results_latencies}
\end{subfigure}
\begin{subfigure}[b]{.49\columnwidth}
  \centering
  \includegraphics[width=\textwidth]{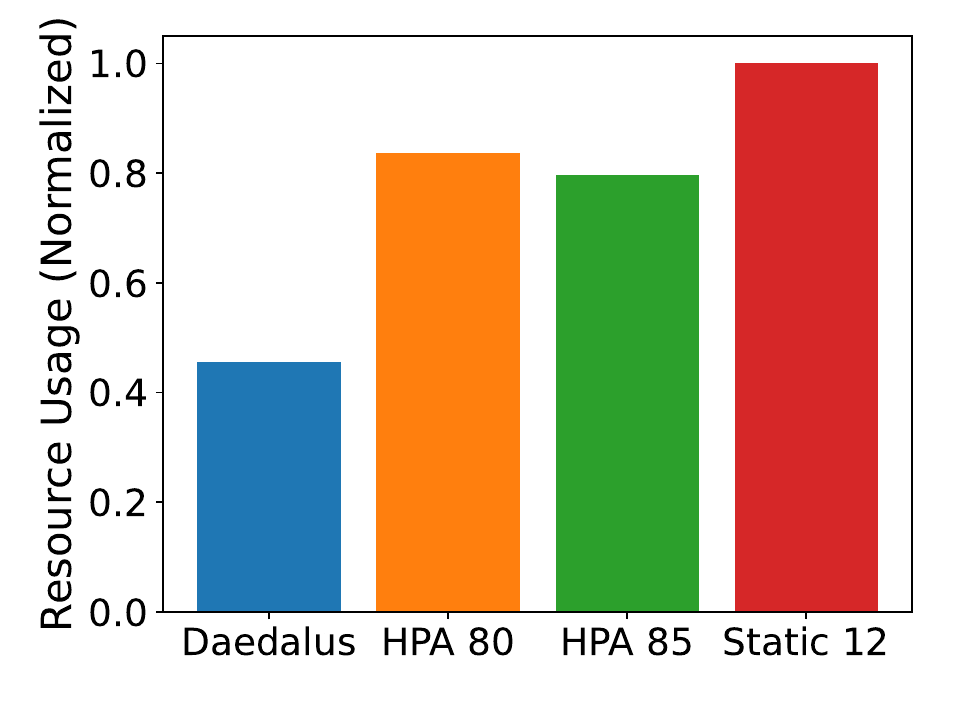}
  \caption{Resource usage}
  \label{fig:ysb_results_resource_usage}
\end{subfigure}
\caption[Yahoo Streaming Benchmark results]{Yahoo Streaming Benchmark results}
\label{fig:ysb_results}
\end{figure}

The results for the Yahoo Streaming Benchmark are shown in~\autoref{fig:ysb_results}. Here, the HPA 80 and 85 deployments scale very similarly. As can be seen in~\autoref{fig:ysb_results_parallelism}, they also allocate more workers than is necessary for a significant portion of the experiment. Both HPA deployments scale over 12 when the workload is around half of its maximum. Daedalus also over-provisions resources during the highest peak due to TSF predictions that the workload will continue to rapidly increase. As with the WordCount experiment, Daedalus scales in more quickly.

As seen in~\autoref{fig:ysb_results_latencies}, Daedalus was proportionally able to achieve the lowest latencies while having slightly more downtime than the HPA deployments. Overall, average latencies for all approaches were similar, being within 1.5 seconds of each other. On average, Daedalus had 9,106 ms, HPA 80 had 7,862 ms, HPA 85 had 8,042 ms, and Static 12 had 7,576 ms. Also for this experiment, the latter did not proportionally achieve lower latencies. The highest latencies for the static scale-out come from when the workload is lowest.

On average, Daedalus used 5.5 workers, HPA 80 used 10, HPA 85 used 9.6, and the static scale-out naturally used 12. Daedalus used 54\% less resources than the static scale-out, 45\% less resources than the HPA 80, and 43\% less resources than the HPA 85. Again, Daedalus was able to process all tuples with reasonable latencies using minimal resources.

\subsubsection{Traffic Monitoring Results}

\begin{figure}
\centering
\begin{subfigure}[b]{.49\columnwidth}
  \centering
  \includegraphics[width=\textwidth]{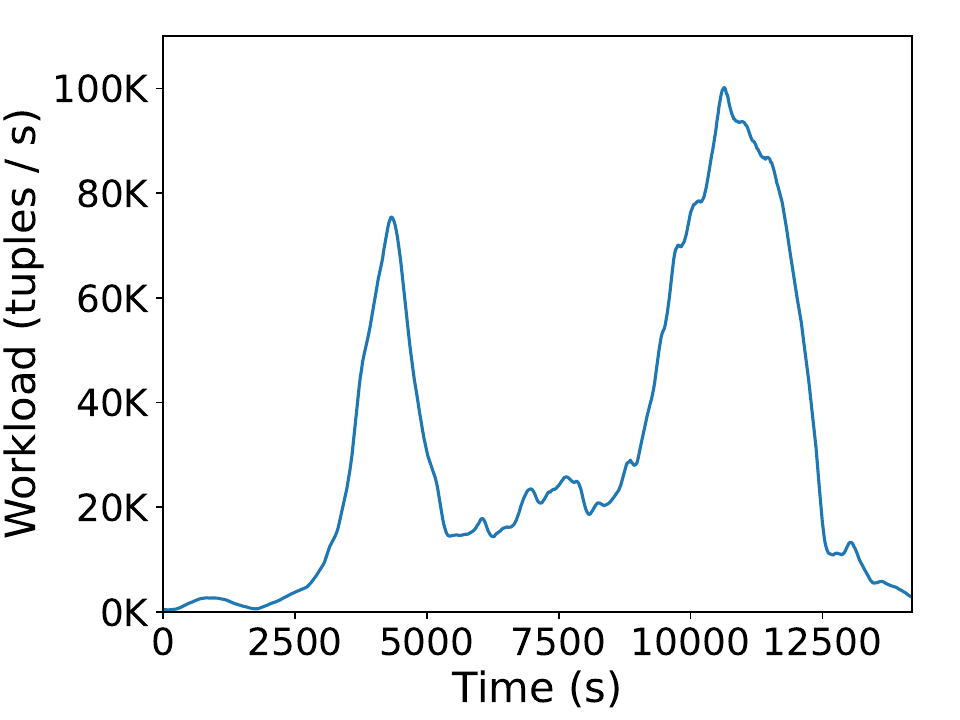}
  \caption{Workload}
  \label{fig:tm_results_workload}
\end{subfigure}
\begin{subfigure}[b]{.49\columnwidth}
  \centering
  \includegraphics[width=\textwidth]{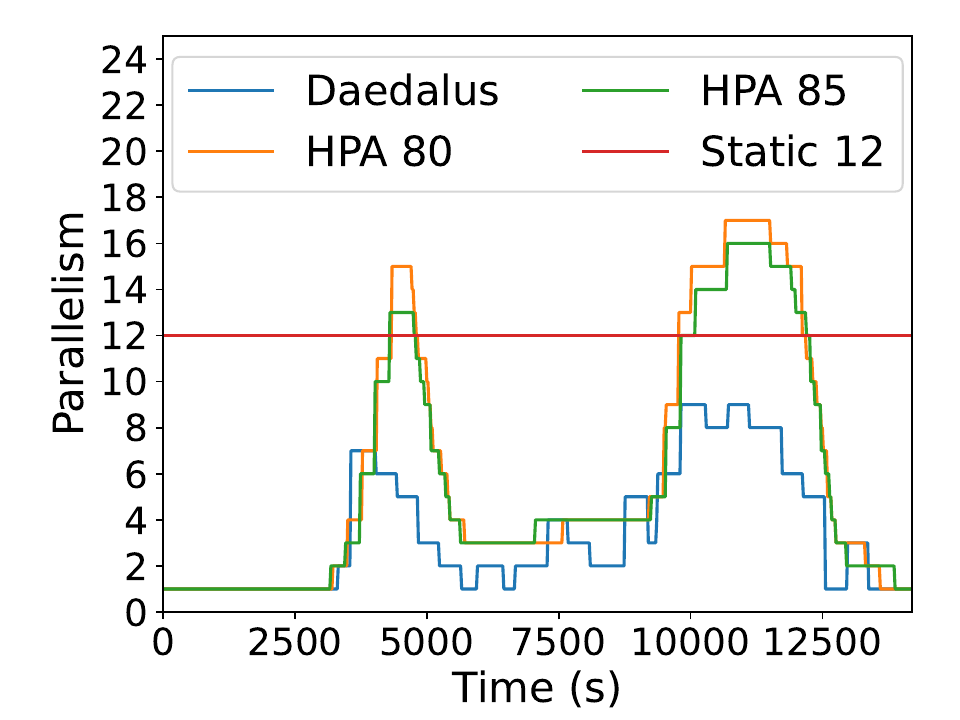}
  \caption{Number of workers}
  \label{fig:tm_results_parallelism}
\end{subfigure}
\begin{subfigure}[b]{.49\columnwidth}
  \centering
  \includegraphics[width=\textwidth]{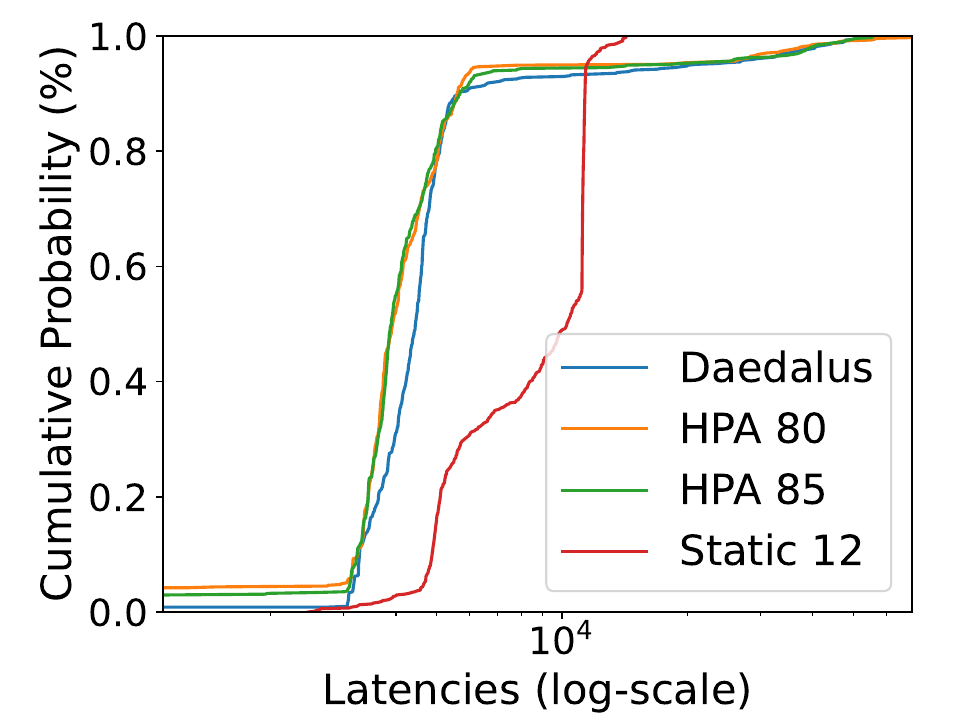}
  \caption{Latency distribution}
  \label{fig:tm_results_latencies}
\end{subfigure}
\begin{subfigure}[b]{.49\columnwidth}
  \centering
  \includegraphics[width=\textwidth]{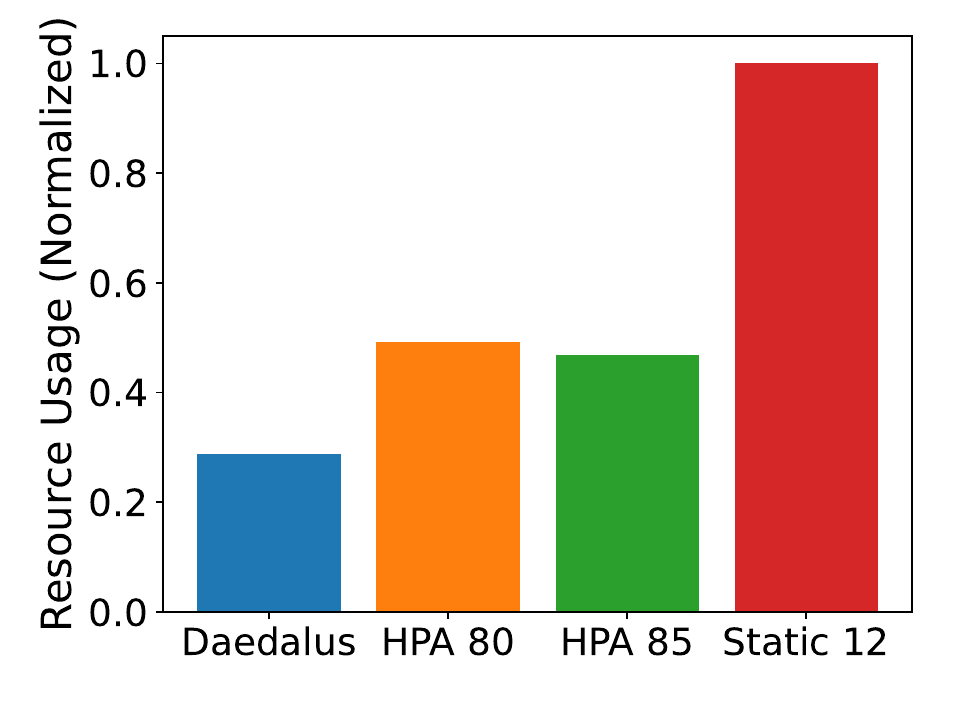}
  \caption{Resource usage}
  \label{fig:tm_results_resource_usage}
\end{subfigure}
\caption{Traffic Monitoring results}
\label{fig:tm-results}
\end{figure}

The results of the Traffic Monitoring experiment can be seen in Figure~\ref{fig:tm-results}. The major challenge of this workload comes from two large spikes where the workload rapidly increases and decreases. As with the Yahoo Streaming Benchmark, the HPA methods scaled similarly and allocated more workers than necessary. Again, Daedalus was able to scale to match the workload. It was able to react more quickly to the falling workload, scaling in faster than both the HPA approaches.

As seen in Figure~\ref{fig:tm_results_latencies}, for the majority of the job, Daedalus and the HPA methods had very similar proportional latencies. All autoscaling approaches had lower average latencies than the static scale-out: Daedalus with 6,176 ms, HPA 80 with 5,566 ms, HPA 85 with 5,671 ms, and Static 12 with 8,778 ms. As with the Yahoo Streaming Benchmark, the lowest latencies for the static scale-out occurred during the highest peaks of the workload.

Daedalus also used fewer resources than the comparison approaches. On average, Daedalus used 3.5 workers, HPA 80 used 5.9, HPA 85 used 5.6, and the static scale-out naturally used 12. Percentually, Daedalus used 71\% less resources than the static scale-out, 41\% less resources than the HPA 80, and 38\% less resources than the HPA 85. For all Flink experiments, Daedalus was able to achieve similar latencies with fewer resources.

\subsection{Kafka Streams Experiment Results}

\begin{figure}
\centering
\begin{subfigure}[b]{.49\columnwidth}
  \centering
  \includegraphics[width=\textwidth]{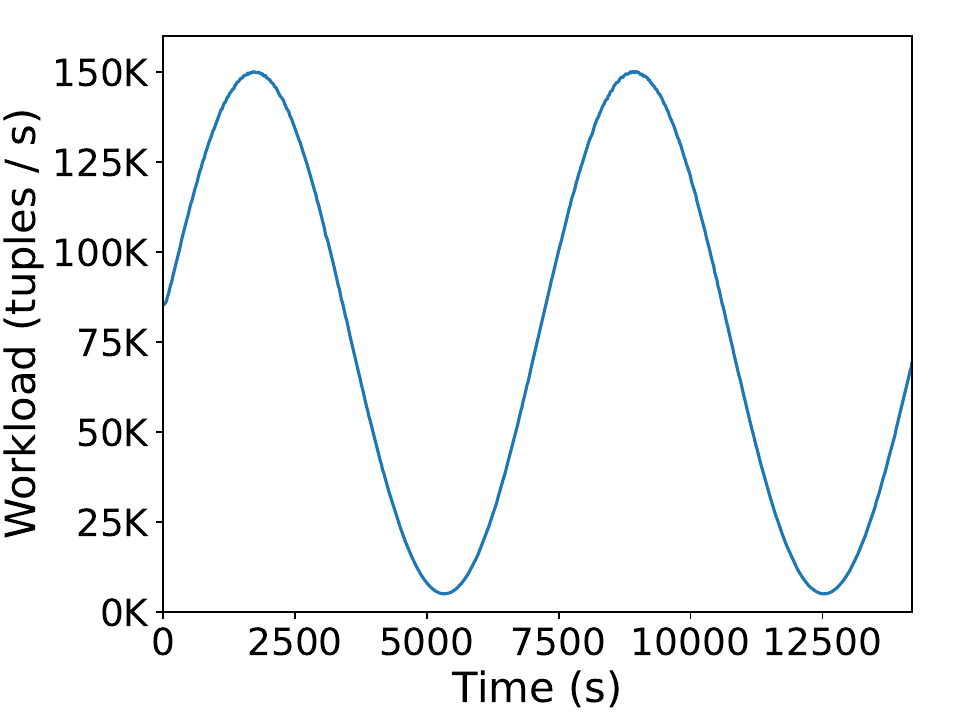}
  \caption{Workload}
  \label{fig:kafka_results_workload}
\end{subfigure}
\begin{subfigure}[b]{.49\columnwidth}
  \centering
  \includegraphics[width=\textwidth]{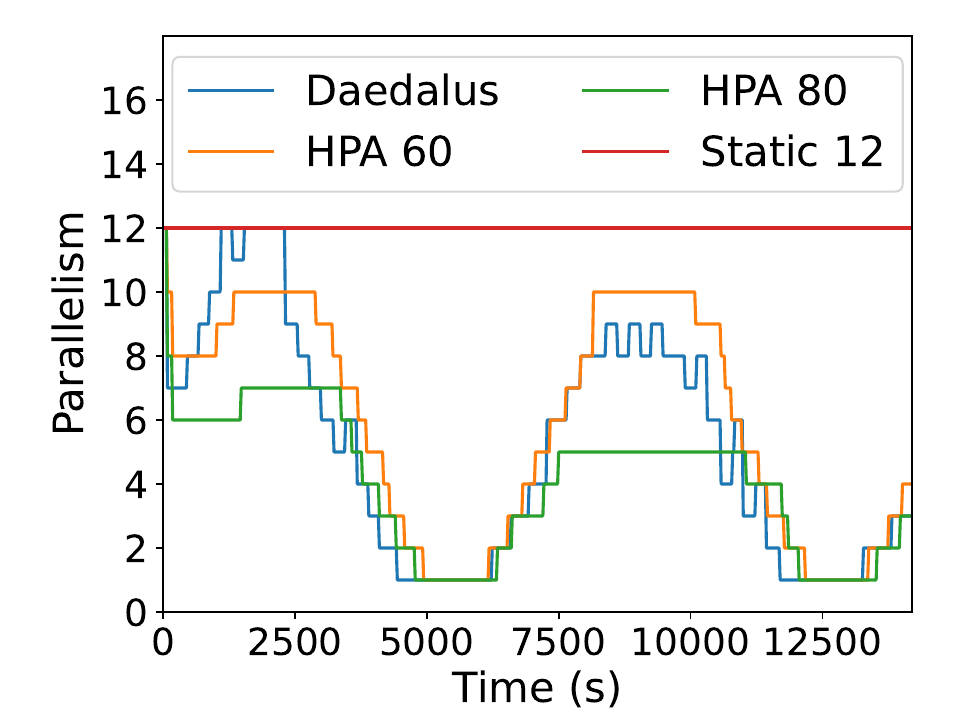}
  \caption{Number of workers}
  \label{fig:kafka_results_parallelism}
\end{subfigure}
\begin{subfigure}[b]{.49\columnwidth}
  \centering
  \includegraphics[width=\textwidth]{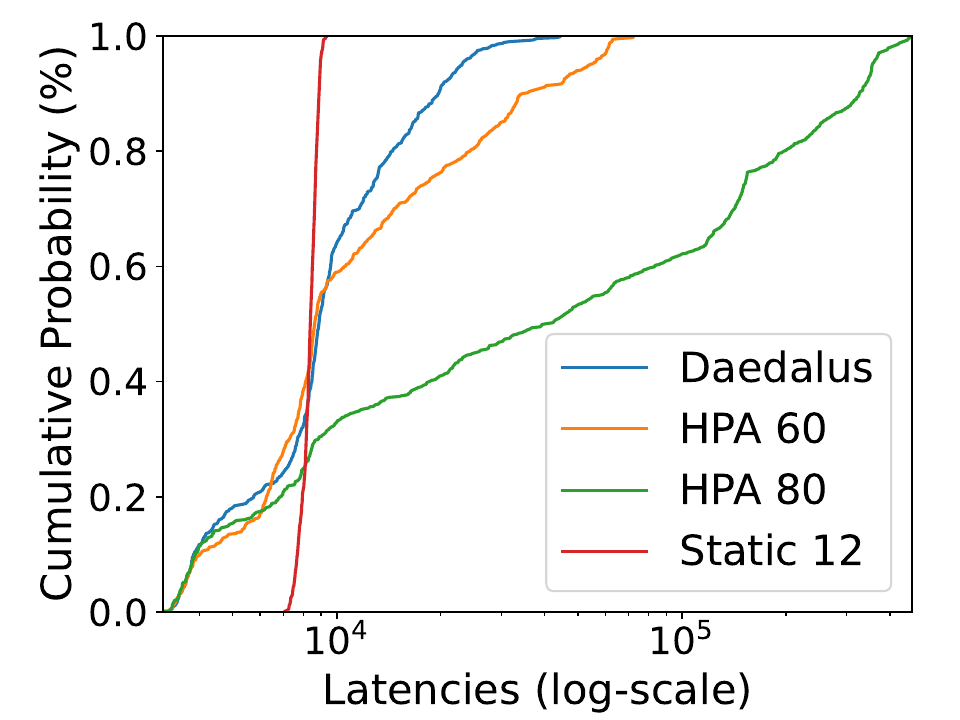}
  \caption{Latency distribution}
  \label{fig:kafka_results_latencies}
\end{subfigure}
\begin{subfigure}[b]{.49\columnwidth}
  \centering
  \includegraphics[width=\textwidth]{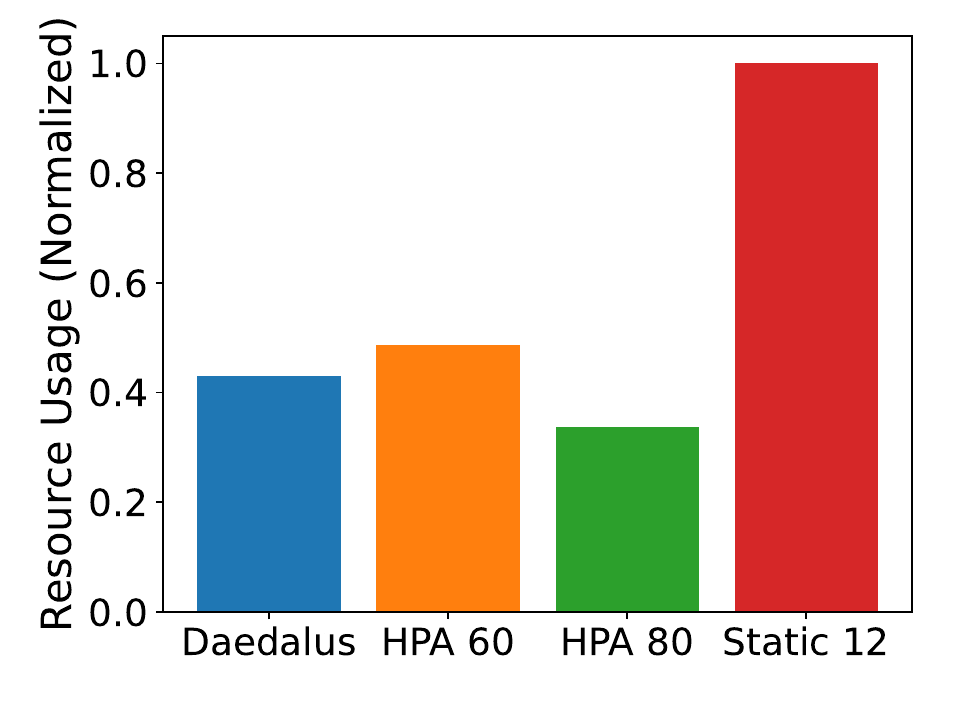}
  \caption{Resource usage}
  \label{fig:kafka_results_resource_usage}
\end{subfigure}
\caption[Kafka Streams WordCount results]{Kafka Streams WordCount results}
\label{fig:kafka_wc__results}
\end{figure}

To show that Daedalus is a general approach that can work with any DSP framework, it is tested with Kafka Streams using the WordCount job. 
The other two DSP jobs used so far, namely Yahoo Streaming Benchmark and Traffic Monitoring, do not qualify for this comparison as no implementations for Kafka Streams exist.
The results are shown in~\autoref{fig:kafka_wc__results}. As with the Flink WordCount job, the job is susceptible to data skew and the maximum capacity at a given parallelism is highly dependent on how data is split among workers. This is especially apparent when observing the peaks of the workload in~\autoref{fig:kafka_results_parallelism}. Unlike in the Flink experiments, HPA 80 was not able to process tuples in a timely manner and under-provisioned resources. This is also evident in the empirical cumulative distribution function. 

The static workload had an almost constant latency and was able to achieve the best overall latencies with an average of 8,343 ms. Daedalus performed next best with an average of 10,566 ms. HPA 60 was slightly worse with an average latency of 15,453 ms. 
Lastly, the HPA 80, which was not able to match the workload had an average of 102,153 ms.  

Compared to the deployments that were able to process the workload, Daedalus used fewer resources. 
On average, Daedalus used 5.2 workers, HPA 60 used 5.8, 
HPA 80 used 4, and the static scale-out naturally used 12. Daedalus used 57\% less resources than the static scale-out and 11\% less resources than the HPA 60. From these results, one can infer that Daedalus is a generally applicable solution. With Kafka Streams, it was able to provide a stable level of service using minimal resources.

\subsection{Comparison with Phoebe}
\begin{figure}
\centering
\begin{subfigure}[b]{.49\columnwidth}
  \centering
  \includegraphics[width=\textwidth]{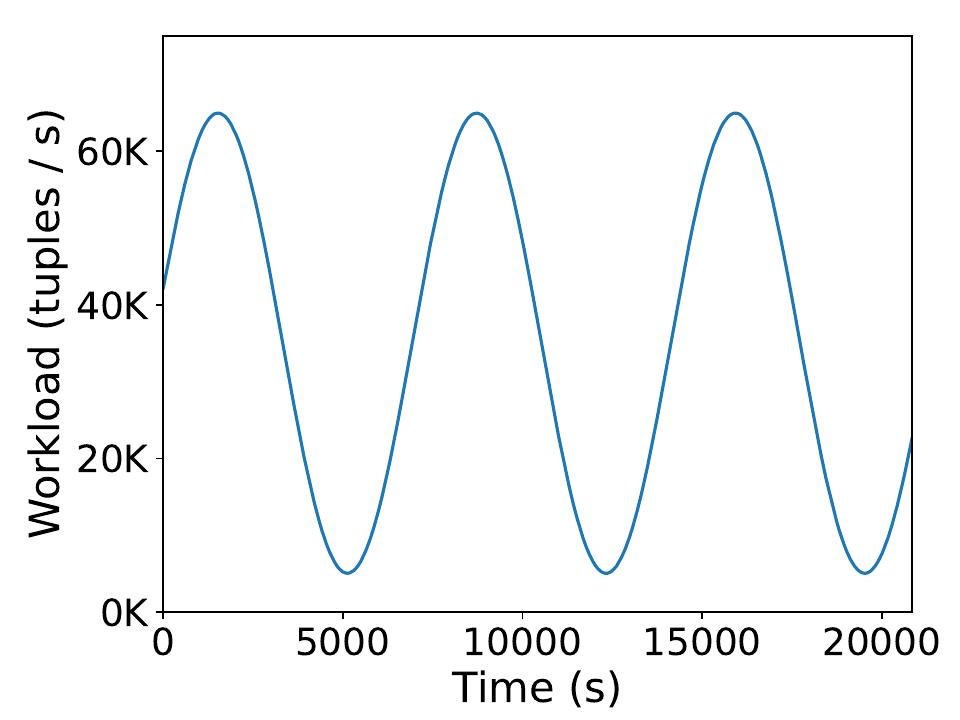}
  \caption{Workload}
  \label{fig:phoebe_results_workload}
\end{subfigure}
\begin{subfigure}[b]{.49\columnwidth}
  \centering
  \includegraphics[width=\textwidth]{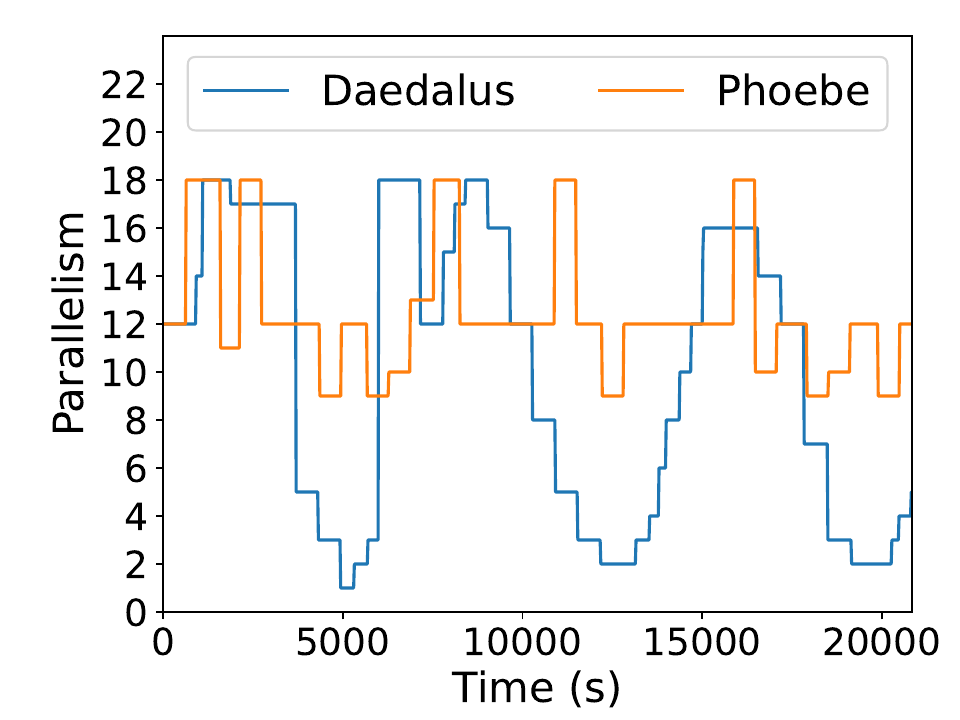}
  \caption{Number of workers}
  \label{fig:phoebe_results_parallelism}
\end{subfigure}
\begin{subfigure}[b]{.49\columnwidth}
  \centering
  \includegraphics[width=\textwidth]{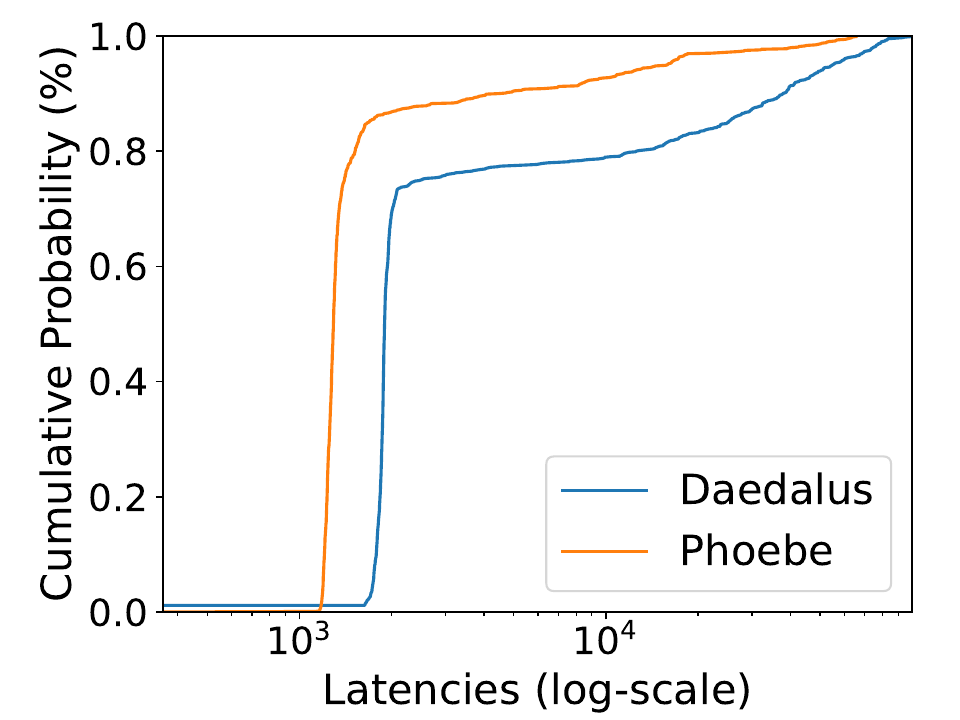}
  \caption{Latency distribution}
  \label{fig:phoebe_results_latencies}
\end{subfigure}
\begin{subfigure}[b]{.49\columnwidth}
  \centering
  \includegraphics[width=\textwidth]{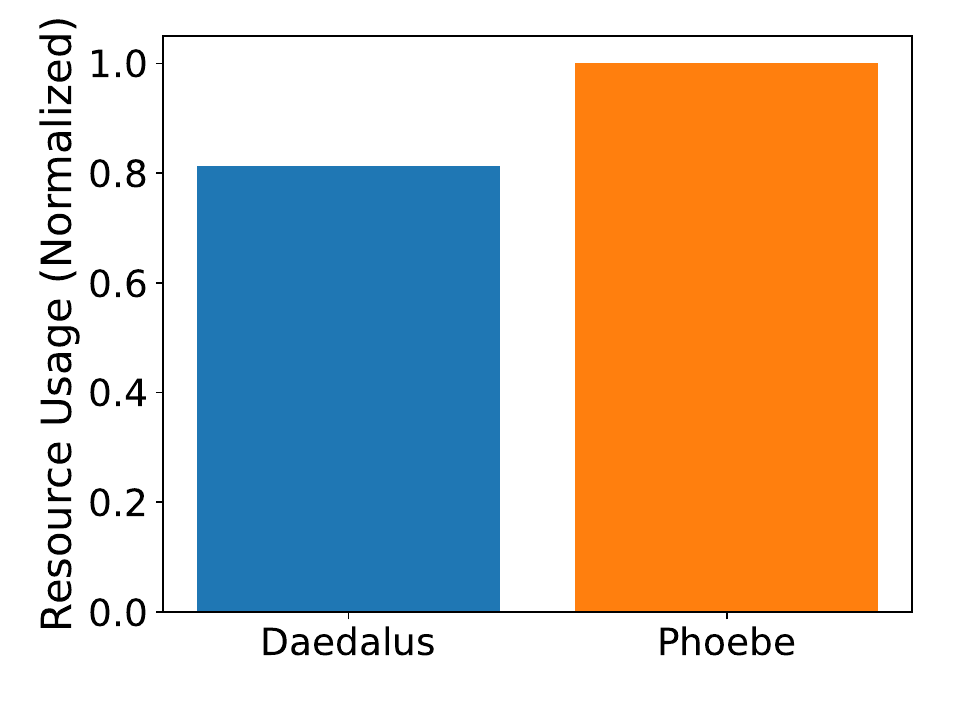}
  \caption{Resource usage}
  \label{fig:phoebe_results_resources}
\end{subfigure}
\caption[Comparison with Phoebe]{Comparison with Phoebe}
\label{fig:phoebe_results}
\end{figure}

The results of comparing Daedalus to Phoebe using the Yahoo Streaming Benchmark can be seen in~\autoref{fig:phoebe_results}. For this experiment, a sine workload was chosen to compare Phoebe's scaling decisions to those in its paper. In addition, the recovery time target of 600 seconds was chosen, since lower recovery time targets (e.g. 180 seconds as in the original publication) caused Phoebe to stay primarily at the maximum scale-out of 18 with the tested workload. 

When looking at the parallelism in~\autoref{fig:phoebe_results_parallelism}, the scaling decisions of Phoebe do not appear to mirror the workload. However, when examining the logs, the scaling decisions are reasonable and balance achieving a minimum latency with a recovery time below the 600 second target. It should also be noted that these results do not match those from the initial paper, where the number of workers were more in line with the workload. In contrast, Daedalus scaled more frequently, but also more in line with the workload. 

When examining latencies, Phoebe outperformed Daedalus by achieving proportionally lower latencies and faster recovery times. Daedalus achieved an average latency of 9,624 ms and a maximum latency of 88 seconds, while Phoebe achieved an average latency of 3,340 ms and a maximum latency of 65 seconds. The maximum latencies indicate the longest time that the system was unavailable, and both maximum latencies were under the target recovery time. %

When comparing resources used during the autoscaling part of the experiment, Daedalus used 19\% less resources, using an average of 10.1 workers while Phoebe used an average of 12.4 workers. However, Phoebe also requires initial profiling runs to build performance models. When incorporating profiling time, Daedalus used 53\% less resources.

\subsection{Discussion}

The conducted experiments show that Daedalus achieves its goals: It allocates sufficient resources to process the incoming workload, processes tuples in a timely manner to achieve reasonable latencies, minimizes resource usage, and makes long-lived scaling decisions. 

For methods such as HPA, adequate thresholds must first be determined. Even then, these fixed thresholds do not guarantee optimal resource usage or meeting QoS requirements. 
As an example, for Flink, the HPA 80 over-provisioned resources (i.e. competitive latencies in exchange for higher resource usage), while for Kafka Streams, it under-provisioned resources (i.e. lower resource usage in exchange for undesirable latencies).
Seemingly, technical variations in DSP systems and the implementation of DSP jobs pose challenges to the HPA methods to achieve generalization.
In addition, the HPA methods do not incorporate data skew, which occurred in these experiments, leading to suboptimal scaling decisions. 

Comparing Daedalus to an approach like Phoebe highlights the trade-offs that can be made in autoscaling. By using initial profiling runs to build latency models, Phoebe can target scale-outs that result in minimal latencies.
However, this comes at the cost of increased resource usage as a result of the required initial profiling runs. In contrast, Daedalus focuses on minimizing resource usage and builds its capacity models by monitoring a running job. It focuses on ensuring throughput and does not explicitly model latency. The autoscaling approach to use ultimately depends on the requirements of the DSP job.
An approach similar to Phoebe is more appropriate when the primary goal is to minimize latencies as much as possible, while Daedalus is more suitable for optimizing resource efficiency.
To further attain lower latencies, Phoebe manually creates a checkpoint before rescaling, minimizing the amount of tuples that need to be reprocessed. In contrast, Daedalus uses Flink's reactive mode, which restarts the job from the last completed checkpoint. While this adaptation could improve achievable latencies, the implementation to configure a manual checkpoint before rescaling is dependent on the DSP system and would make Daedalus less generalizable.

The effectiveness of Daedalus relies on its ability to accurately ascertain worker capacity across all scale-outs. While it is difficult to truly determine the maximum capacity given factors like data-skew, it can be approximated by ensuring ample tuples in the data source to saturate the system and observing the throughput at different scale-outs. Comparing the observed capacities to the estimates from Daedalus gives insights into the estimation accuracy. Generally, the estimated capacities typically differ less than 5\% from the observed capacities, with the majority between 0\% and 3\%. Therefore, it can be concluded that Daedalus can accurately estimate capacity.

Since the quality of scaling decisions are also impacted by TSF, the accuracy of TSF predictions was evaluated. TSF predictions were generally accurate with errors typically falling below 5\%. In fact, the threshold for poor predictions at 25\% was never reached. 

In this evaluation, the recovery time heuristic was tested using the time needed to recover from a rescale. However, real failures typically incur longer recovery times since the DSP system must first detect the failure. While Daedalus accounts for failure by using a worst-case recovery time calculation, an evaluation that injects failures is left for future work.
In general, a lower desired recovery time will lead to higher resource utilization, making recovery time the primary factor influencing autoscaling decisions. 
On the contrary, a higher desired recovery time will have less impact on scaling decisions, with actual processing capacity becoming the key determinant. 
To align with HPAs for comparison purposes, we opted for a higher recovery time of 600 seconds without exploring the boundaries or quantifying the precise influence of the recovery time parameter.
In our experiments, the predicted recovery time was almost always greater than the measured recovery time. However, due to the worst-case calculation, the accuracy ranges wildly from a 1\% difference to a 140\% difference when comparing the actual and predicted recovery times. 
This lack of precision is a limitation to this approach and could be improved in future work.

\section{Conclusion}
\label{sec:conclusion}

This paper presents Daedalus, a self-adaptive autoscaling approach for DSP systems. 
Daedalus monitors running DSP jobs, builds worker-level capacity models using readily available CPU and throughput metrics, and scales resources to meet adaptation goals. 
Its primary objectives are to process incoming workloads efficiently while minimizing resource usage and to make long-lived scaling decisions to reduce system downtime.
After evaluation with three representative DSP jobs in two DSP systems, Flink and Kafka Streams, Daedalus proves to be an effective autoscaling method. 
It accurately estimates the maximum processing capacity across different scale-outs and matches resources to the incoming workload. 
Using up to 71\% fewer resources, Daedalus achieves latencies comparable to those of a static scale-out and HPA. 
When compared to Phoebe, a state-of-the-art approach that explicitly models latency, Daedalus does not quite achieve the same low latencies, nonetheless, it still provides a stable level of service with fewer resources than Phoebe. 
A current limitation of Daedalus is its inability to target scale-outs for minimal latencies, and incorporating a fitting latency model is considered a direction for our future work.

\bibliographystyle{ACM-Reference-Format}
\bibliography{references} 

\end{document}